# RESEARCH ARTICLE

# London's Blue Light Collaboration Evaluation: A Comparative Analysis of Spatio-temporal Patterns on Emergency Services by London Ambulance Service and London Fire Brigade

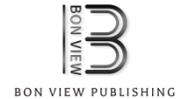

**Fangyuan Li, Yijing Li\* and Luke Edward Rogerson**
*1 Department of Geography, King's College London, UK.*
*2 Department of Informatics, King's College London, UK.*
*3 London Ambulance Service NHS Trust, UK.*

**\*Corresponding author:** Yijing Li, Department of Informatics, King's College London, UK. Email: yijing.li@kcl.ac.uk

**Abstract:** With rising demand for emergency services, the London Ambulance Service (LAS) and the London Fire Brigade (LFB) face growing challenges in resource coordination. This study investigates the temporal and spatial similarities in their service demands to assess potential for routine cross-agency collaboration. Time series analysis revealed aligned demand peaks in summer, on Fridays, during daytime hours, and were highly sensitive to high-temperature weather conditions. Bivariate mapping and Moran's I indicated significant spatial overlaps in central London and Hillingdon. Geographically Weighted Regression (GWR) examined the influence of socioeconomic factors, while Comap analysis uncovered spatiotemporal heterogeneity across fire service types. The findings highlight opportunities for targeted collaboration in high-overlap areas and peak periods, offering practical insights to enhance emergency service resilience and efficiency.

**Keywords:** emergency response; demand patterns; London ambulance service; London fire brigade; urbanization; blue light collaboration; spatiotemporal patterns

## 1. Introduction

The London Ambulance Service (LAS) and the London Fire Brigade (LFB) are vital components of London's emergency response system, critical in safeguarding public health and safety. However, in recent years, emergency service agencies have been facing mounting pressures. In 2023 alone, LAS received approximately 2.28 million emergency calls—a 9% increase compared to 2.10 million in 2021. Similarly, the LFB responded to 126,036 incidents in 2023, reflecting a 15.6% rise from around 109,000 incidents in 2021 [1, 2]. With population growth, increasing urban complexity, and intensifying resource pressures, the need to coordinate emergency service deployment and resource allocation to meet surging demand has become an urgent issue.

Cross-agency collaboration has been identified as a potential response strategy. During the COVID-19 pandemic, cross-agency collaboration significantly enhanced the overall capacity to respond to the surge in emergency demand. For instance, the LFB deployed over 300 firefighters to assist with ambulance driving and patient transfers, while nearly 100 firefighters joined multi-agency response teams supporting LAS. These practices effectively strengthened emergency service coordination and the continuity of critical services.

At the policy level, since 2015, the UK government has initiated a consultative process to deeply explore the collaborative capabilities among emergency service sectors, including the police, fire, and ambulance services, which was named as Blue Light Collaboration. The goal was to promote a more interconnected approach. The consultation document specified that to meet the demands of these three emergency services, strengthening collaboration and driving innovation were imperative [3]. Furthermore, the Policing and Crime Act 2017 granted Police and Crime Commissioners authority over the governance of fire and rescue services, mandating emergency service agencies to uphold a "high-level duty to collaborate.[4]"

Despite established policies and practical experiences, existing cross-agency collaborations have ~~largely~~ mostly been reactive, triggered by crisis events. Routine collaboration models in day-to-day emergency services remain underexplored. In particular, understanding of the spatio-temporal overlaps in fire and ambulance service demands remains limited, which could offer new opportunities for optimising resource allocation and improving collaborative efficiency.

Based on these findings, this application research aims to examine the temporal and spatial similarities in service demand patterns between LAS and LFB. It seeks to identify opportunities for routine collaboration and to provide empirical evidence to support normalised cross-agency coordination. The research objectives are as follows: (1) to analyse the periodic patterns of fire and ambulance service demands over time and their associations with weather conditions; (2) to examine the spatial overlap of the two types of service demand, identify high-demand areas, and compare their relationships with socioeconomic factors and the spatial heterogeneity of these relationships; (3) to identify the spatiotemporal heterogeneity characteristics of different types of fire service demands.

## 2. Literature Review

### 2.1. Interoperability in Multi-agency Emergency Response

The UK government has consistently emphasised establishing collaborative frameworks among emergency service agencies and integrating these into routine training and multi-agency incident management. The Civil Contingencies Act 2004 formally introduced legal requirements demanding cooperation among emergency services [5]. To further this initiative, the government introduced the Joint Emergency Services Interoperability Principles (JESIP) in 2012, defining interoperability as "the extent to which organisations can work together coherently as a matter of routine" [6, 7]. JESIP proposed five core principles—co-locate, communicate, coordinate, jointly understand risk, and share situational awareness—to standardise inter-agency communication, task allocation, and decision-making processes, enhancing practical interoperability [6].

JESIP played a critical role during the early stages of the COVID-19 pandemic. For example, the Kent and Medway Local Resilience Forum (LRF) established a Multi-Agency Information Cell (MAIC), integrating resources from local government, police, and fire services to provide a unified platform for information sharing and decision-making. Daily briefings and data modelling conducted by the MAIC improved situational awareness, resource allocation, and public communication strategies. This approach was subsequently replicated in other regions to address varying pandemic phases and challenges [8].

However, significant challenges persist regarding the application of interoperability frameworks in complex real-world scenarios. The official report on the Manchester Arena attack highlighted considerable shortcomings in JESIP implementation, specifically a failure in information sharing among emergency services, resulting in the Greater Manchester Fire and Rescue Service (GMFRS) arriving at the scene more than two hours after the incident [9]. Interoperability challenges extend beyond the UK, as emergency services worldwide frequently encounter difficulties [10, 11].

Current research on emergency service collaboration primarily addresses the problems in social mechanisms, organisational structures, and cultural differences. For instance, studies by the Emergency Services Collaboration Working Group indicate that collaboration within the UK remains inconsistent, and restricted by local politics, funding disparities, and inter-organisational cultural differences. These studies argue that beyond legal and procedural frameworks, formulating a culture of trust and shared understanding among agencies is urgently required [12, 13]. Similarly, research by Wankhade highlights the fragmented governance structures within the UK's emergency response system, noting the absence of unified management and command structures [14]. In contrast to these socially oriented studies, Neville and colleagues suggest incorporating data-driven methodologies, such as spatial modelling and causal analysis, into emergency collaboration frameworks. They highlight the significance of developing scenario-based integrated models and iteratively testing these tools under real-world conditions to enhance interoperability outcomes [15].

### 2.2. Spatio-temporal Demand Patterns in Emergency Services

Numerous case studies have illustrated that the spatiotemporal modelling and data integration capabilities provided by Geographic Information Technologies significantly support efficient collaboration and information sharing among emergency services [16, 17]. Currently, several studies have analysed demand patterns for individual emergency services and proposed targeted preventive strategies.

Firstly, in policing, spatiotemporal analysis facilitates the precise allocation of police resources to micro-level hotspots, illustrating the concept of "hot spot policing." Extensive empirical research has confirmed that hotspot policing strategies derived from spatiotemporal modelling have yielded notable results [18, 19].

Additionally, as essential emergency response agencies in London, numerous studies have investigated the spatiotemporal trends of ambulance and fire service demands to enhance resource allocation and deployment efficiency. Existing research on ambulance demand highlights (1) a significant spa tial clustering in the workload distribution [4, 20]; and (2) a distinct temporal pattern, featuring by noticeable seasonal variations, particularly in response to extreme weather conditions [21]. Thornes et al. [46], for example, analysed ambulance call data in Birmingham and found increases in call volumes during cold weather. Similarly, Turner et al. [47] demonstrated that high temperatures significantly contributed to ambulance demand for cardiovascular emergencies in Brisbane.

Similarly, empirical studies on fire department demand have also shown that fire incidents also demonstrate seasonal variation in time series [23, 24], with distinct spatial clustering patterns [24, 25]. Meteorological factors also exert considerable influence on fire-related service demand. For instance, Corcoran found that secondary fires increase substantially during dry conditions, while residential fire rates decrease in high-humidity environments. Moreover, extreme temperature conditions elevate fire risks [49]. Gill's study in Kakadu National Park confirmed increased fire probability under high temperatures and low dew points [50].

Beyond environmental drivers, socioeconomic characteristics have also been shown to correlate with emergency service demand. For fire services, several studies link increased fire risk to social deprivation, poor housing, or low homeownership [24, 49, 57]. For ambulance services, factors such as population ageing, poor health conditions, limited access to private transport, and low educational attainment are frequently cited. Kawakami [54] found that elderly populations and low car ownership increased non-urgent ambulance calls in Japan. Wong [55] demonstrated the role of education and urban-rural disparities in ambulance usage, and Earnest [56] highlighted household size as a key driver of ambulance call volumes in Singapore.

Although extensive research has examined fire and ambulance service demands individually, research on the spatiotemporal correlations between the two types of services remains limited. A few studies have attempted to bridge this gap. For example, Wuschke investigated the spatiotemporal patterns of residential burglaries and structural fires in Surrey, British Columbia, demonstrating that both types of incidents exhibited clustering in space and time [26]. Similarly, Clare analysed the spatial and temporal intersections among police, fire, and ambulance incidents in Surrey, Canada, between 2011 and 2013. By dividing the area into spatial cells of varying sizes and recording incident counts over the entire study period, annually, and monthly, Clare's study confirmed substantial spatial and temporal overlap across emergency service incidents. Based on these findings, the study proposed a classification framework to help emergency services identify priority areas for collaboration and to facilitate proactive intervention and resource integration [18].

However, most existing studies continue to apply traditional mapping and clustering techniques independently to each emergency service, without exploring the spatial interrelationships between them. Integrated visual analytics that reveal the joint spatial dynamics of multiple services remain rare, and few studies utilise robust statistical methods—such as bivariate spatial autocorrelation—to assess the significance of cross-service clustering. Comparative analyses of shared or divergent influencing factors also remain underdeveloped. As a result, cross-agency coordination frameworks are often descriptive in nature, lacking deeper strategic insight.

To address these gaps, this study investigates not only the spatiotemporal overlap between LAS and LFB demand patterns, but also examines the similarity in their external drivers, including weather conditions and socioeconomic characteristics. This dual approach offers novel insights into the common underlying mechanisms of emergency service pressures and provides an evidence base for identifying priority zones for routine cross-agency collaboration.

## 3. Research Methodology

### 3.1. Data

(1) The LAS demand data were obtained from the London Ambulance Service, part of the NHS system, covering the period from 2011 to March 2023. Two datasets were included. The first dataset, LAS Incidents, recorded monthly counts of 999 call incidents at the LSOA level. The second dataset, LAS Hourly Call, provided hourly counts of 999 calls across the entirety of London, without geographic detail. It is important to acknowledge that the limited spatial and temporal granularity of the LAS datasets, along with inconsistencies between them, constrained detailed spatiotemporal characterisation of demand dynamics.

(2) The LFB incident data were sourced from the London Datastore and included records of LFB incidents from 2009 [27]. Each record contained an incident number, precise geographical coordinates (latitude and longitude), exact timestamp (to the second), and incident type (e.g., fire, special service, false alarm). Some records lacked precise location but retained a "rounded location." Validation tests indicated that the average spatial error between precise and rounded locations was approximately 20 metres. Therefore, when precise coordinates were unavailable, approximate locations were used as substitutes, which may have introduced minor spatial inaccuracies.

(3) Weather data were obtained from the Weather Sparks website and covered daily meteorological observations from 2011 to 2023 [28]. Measurements were recorded at the Heathrow Airport Meteorological Station. Variables included date, average temperature, average dew point, average wind speed, and wind direction.

(4) Socio-economic data were compiled from multiple official sources. The primary source was Nomis, the official census and labour market statistics platform, based on the 2021 Census (conducted on 21 March 2021) [29]. This dataset covered demographic variables such as age, education level, health status, vehicle ownership, household size, and homeownership rates. Additionally, supplementary datasets were incorporated, including road casualty data from London Datastore (2018) [30], the UK Government's Index of Multiple Deprivation (IMD) published in 2019 [31], and the Public Transport Accessibility Levels (PTAL) index released by Transport for London in 2015 [32]. It should be noted that some temporal mismatches existed between the socio-economic datasets and the service demand data in this study (2018–2023), which may have introduced slight time-related biases.

### 3.2. Methods

**Figure 1**
**Workflow of Analysis**

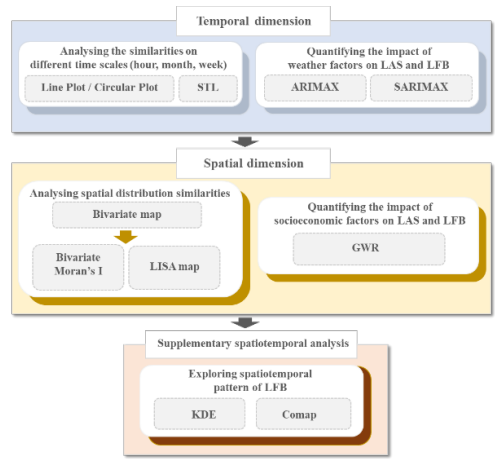

### 3.2.1. Research design

The research is designed with three steps. Firstly, temporal sequence analysis to identify similarities between demands for LFB and LAS. This includes using line graphs and circular diagrams to visualise demand trends, applying STL decomposition to detect seasonal patterns, and employing SARIMAX models to examine the influence of weather factors on service demand. Secondly, spatial analysis investigates the spatial distribution of LFB and LAS demand in assistance of bivariate maps, while statistical validation is conducted using bivariate Moran's I and LISA maps to identify spatial correlations and clustering patterns. Additionally, GWR is applied to explore the influence of demographic and sociopolitical factors on spatial demand patterns. Lastly, the research examines the spatiotemporal patterns of LFB using KDE and comap to visualise and interpret demand dynamics.

### 3.2.2. Time Series Analysis: ARIMAX and SARIMAX

The Autoregressive Integrated Moving Average (ARIMA) model and the Seasonal Autoregressive Integrated Moving Average (SARIMA) model fall within the domain of time series forecasting and analysis. They were initially introduced by Box and Jenkins [34]. The ARIMA model, denoted as $ARIMA(p, d, q)$, is based on the assumption that a given time series is non-stationary. To model such data, a preliminary differencing step is applied to induce stationarity, followed by fitting the $ARMA(p, q)$ model to the newly differenced time series. $(p, d, q)$ represent the AR order $(p)$, differencing order $(d)$, and MA order $(q)$ respectively.

SARIMA, an extension of the ARIMA model, accommodates time series data characterized by seasonal components. SARIMA introduces three additional hyperparameters represented as $(P, D, Q)^S$, which mirror the $(p, d, q)$ structure of ARIMA but specifically cater to the seasonal components.

The SARIMAX and ARIMAX models are extensions of the ARIMA and SARIMA models, respectively, that include the incorporation of exogenous variables denoted as "X." These exogenous variables can be integrated using a multivariate linear regression equation. In the context of this study, these external variables are factors related to weather, such as temperature, dew point, and wind speed. ARIMA and SARIMA models have been widely used to forecast ambulance demand and fire incidents, demonstrating notable effectiveness. Their extended forms, ARIMAX and SARIMAX, incorporate external meteorological variables while addressing non-stationarity and seasonality. In this study, these models were applied to explore the relationship between LAS and LFB service demands and weather conditions.

### 3.2.3. Spatial Visualisation: Bivariate map

Bivariate map origins back to the 1970s, with the initial set of maps being released by the United States Census Bureau [37]. In recent decades, bivariate mapping finds widespread application across various domains of earth science and social research for simultaneous visualisation of two variables [38]. For example, Teuling employed bivariate maps to illustrate the state of temperature and relative humidity [39]. Hall and colleagues combined infant mortality rates with national environmental vulnerability levels to craft bivariate choropleth maps to explore the intersection between environmental factors and poverty [40]. Similarly, Calka generated Bivariate choropleth charts based on land cover intensity and population growth intensity, facilitating the portrayal of relationships between population dynamics and land utilisation [37].

Bivariate mapping displays two variables on a single map by blending their values, rather than presenting them separately [33]. When variables are presented individually, each typically has its own colour legend. For bivariate mapping, we applied a two-dimensional colour legend to clarify spatial relationships between the variables. In Figure 2, the bottom left and top right cells along the diagonal represent the lowest and highest spatial associations for LAS and LFB, respectively.

Bivariate mapping allows for the concurrent representation of two geographic variables within a single spatial unit, offering an effective means to explore spatial co-occurrence and divergence. In this study, it was employed to jointly visualise the spatial distribution of LAS and LFB service demands, thereby enabling a more nuanced assessment of their spatial relationship.

**Figure 2**
**Colour legend explanation for bivariate map**

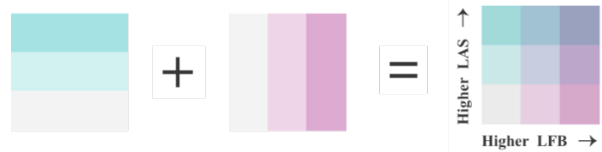

### 3.2.4. Bivariate Moran's I and Bivariate LISA

Spatial autocorrelation refers to the phenomenon where neighbouring variable values demonstrate similarity [35]. Moran's $I$ is a widely used measure of global spatial autocorrelation, calculated as the cross-product between a variable and its spatial lag, based on deviations from the mean. Bivariate Moran's $I$ extends this to assess whether two variables share similar spatial distribution patterns. The formula for Bivariate Moran's $I$ is as follows:

$$I_b = \frac{n}{S_0} \frac{\sum_i \sum_j \omega_{ij} (x_i - \overline{x})(y_j - \overline{y})}{\sqrt{\sum_i (x_i - \overline{x})^2 \sum_j (y_j - \overline{y})^2}}$$
(1)

where $n$ represents the number of spatial units, $S_0$ stands for the sum of weights, $\omega_{ij}$ denotes the spatial weight between locations $i$ and $j$. $x_i$ and $y_i$ correspond to the observed values of variables X and Y at positions $i$ and $j$ respectively.

A positive value of bivariate Moran's I indicates that the two variables exhibit similar spatial clustering patterns, whereas a negative value suggests an inverse spatial relationship. The magnitude of the value reflects the strength of the mutual association. While Moran's I can describe the overall spatial pattern's characteristics, it doesn't pinpoint the specific cluster locations.

To address this limitation, Anselin introduced LISA in 1995 [36], a statistical method for assessing spatial autocorrelation at individual locations. The extension of LISA to bivariate data assesses the similarity of spatial patterns between two variables and identifies regions where the spatial patterns of association between the two variables are statistically significant, which can be computed using the following formula:

$$I_i^{(X,Y)} = \frac{\omega_{ij} (x_i - \overline{x})(y_j - \overline{y})}{\sqrt{\sum_i (x_i - \overline{x})^2 \sum_j (y_j - \overline{y})^2}}$$
(2)

where $\omega_{ij}$ denotes the spatial weight between locations $i$ and $j$. $x_i$ and $y_j$ correspond to the observed values of variables X and Y at positions $i$ and $j$ respectively. This localised approach makes it possible to identify clusters where the spatial association between LAS and LFB demands is statistically significant, offering a deeper understanding of how the two services may spatially co-locate or diverge.

### 3.2.5. Geographically Weighted Regression (GWR)

To analyse the relationship between LAS and LFB service demands and socio-economic variables, and to examine their spatial heterogeneity, this study employed the GWR model. Unlike conventional global regression models, GWR reveals spatially varying relationships across local areas, thereby providing a more comprehensive account of spatial non-stationarity. The GWR model incorporates spatial autocorrelation elements and estimates regression coefficients at each specific location, accommodating the interactions of neighbouring data points. Additionally, it thoroughly considers the impact of geographical location, leading to spatially varying relationships among variables [41].

The formulation of this model can be described as follows:

$$y_i = \beta_0(u_i, v_i) + \sum_{i=1}^{k} \beta_j(u_i, v_i) x_{ik} + e_i$$
(3)

where the variable $y_i$ represents the demand for LAS or LFB, $x_{ik}$ denotes the selected socio-demographic population characteristic variables. $k$ stands for the total number of spatial units (LSOAs), and $e_i$ represents the random error term. The term $(u_i, v_i)$ signifies the spatial location of sample $i$.

### 3.2.6. KDE and comap

To capture the spatiotemporal dynamics of emergency service demand, this study integrated Kernel Density Estimation (KDE) with the Comap technique. KDE was first used to construct spatial distribution maps of LFB service demand across different time periods, calculating the density of events within their respective neighbourhoods to depict the distribution of spatial data. Specifically, KDE computes the density of point attributes around each output raster cell, resulting in a smoothly contoured surface for each point. As a result, it generates a cohesive density surface that spans the entire region. These surfaces reach their highest values at corresponding points and gradually decrease as the distance from the point increases [42].

Subsequently, to visualise the temporal dynamics of these KDE-generated spatial patterns, the Comap technique was adopted. Comap is an exploratory analysis technique that operates on the principle of segmenting data based on time and subsequently presenting the outcomes through a plethora of charts or maps, which are sequentially arranged to accentuate the temporal variations in the distribution of the targeted variables. This method has found extensive utility in investigations concerning temporal and spatial dynamics. For instance, Asgary utilised the comap approach to unveil how structural fire incidents in Toronto exhibit fluctuations across different time segments within a day, days of the week, and months of the year [42]. Similarly, Corcoran leveraged UK South Wales fire data to construct comaps, highlighting its advantages in analysing emergency medical services' temporal and spatial dynamics [23].

Comap employs sets of "small multiples" of diagrams to effectively illustrate alterations in a pair of variables across temporal sequences [43]. Within this methodology, the spatial occurrences of LFB demands (represented by coordinates x and y) are distinctly described across different temporal intervals (z).

The primary concept involves segmenting the original data into temporal subsets containing single or paired variables for observation. For instance, Figure 3 illustrates a univariate depiction, presenting distribution shifts over each day's distinct hourly segments. Similarly, Figure 4 displays a bivariate representation, showcasing spatial distribution influenced by the interaction of various times of day and months.

Subsets should overlap with neighbours and contain roughly equal observations to mitigate classification bias. This approach mitigates classification bias. For instance, non-overlapping hourly partitions might treat events independently, missing broader temporal trends [23, 42].

**Figure 3**
**Univariate comap showing distribution of different time intervals in a day**

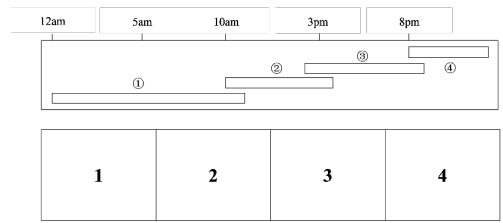

**Figure 4**
**Time interval and monthly intersection bivariate comap**

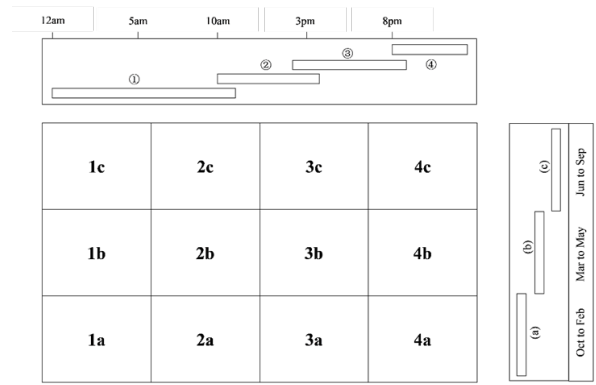

## 4. Results

### 4.1. Temporal patterns analysis of LAS and LFB Demands

#### 4.1.1. Temporal patterns on demands by week, month, day-of-week and hour-of-day

Figure 5 illustrates the average daily call demand for LAS and LFB on weekly basis from January 2011 to March 2023. There were fluctuations for LAS demands without a discernible trend of significant growth or decline pre-pandemic; but the demand experienced pronounced volatility with the onset of the pandemic in early 2020, such as a sharp decrease from March 2020, followed by a surge to peak levels during the winter of 2020, then a decline in February 2021, followed by a recovery in July. In contrast, the demand for the LFB demonstrated relatively stable trends from 2011 to 2023, with no significant fluctuations over the pandemic, but kept its consistently higher demand during summer seasons. Additionally, a significant surge in LFB demand occurred in February 2018 due to the possible impact from flooding disaster.

**Figure 5**
**Weekly average incidents: LFB (Orange) vs LAS (Blue) (Jan 2011-Mar 2023)**

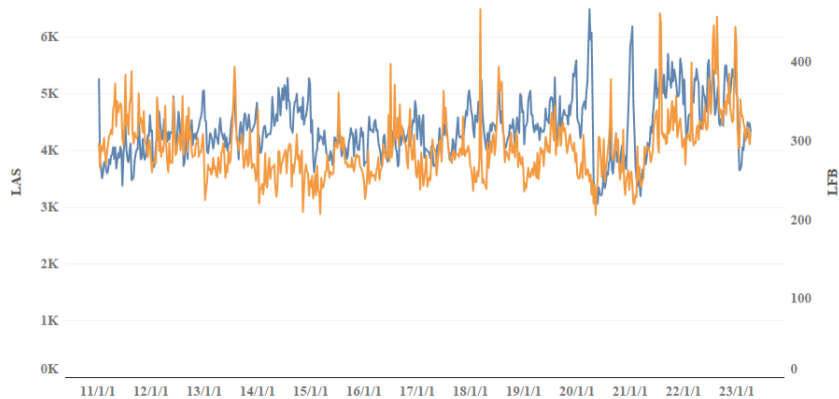

Upon aggregating the data by month, Figure 6 illustrated the average daily call demand for LAS and LFB on monthly basis from 2018 to 2022, with varied symbolised line representing for corresponding year, and the two orange dashed lines indicating 2020 and 2021 which were affected by the COVID-19 pandemic. In non-pandemic years (2018, 2019, and 2022, grey solid lines in Figure 6), the monthly trends for LAS's demand were consistent in general. There was a noticeable increase in demand around March and July, followed by declines in April and September and subsequent growths afterwards. During the pandemic, demands for LAS experienced a sharp increase at the onset of the outbreak in March 2020, followed by a drop to the bottom May, but then a surge to over 5000 calls by December 2020. The demand in February 2021 presented a significant decrease, followed by continuous growth until July and subsequent stability at a relatively high level.

**Figure 6**
**Average Daily LAS and LFB Demand by Month (2018-2022)**

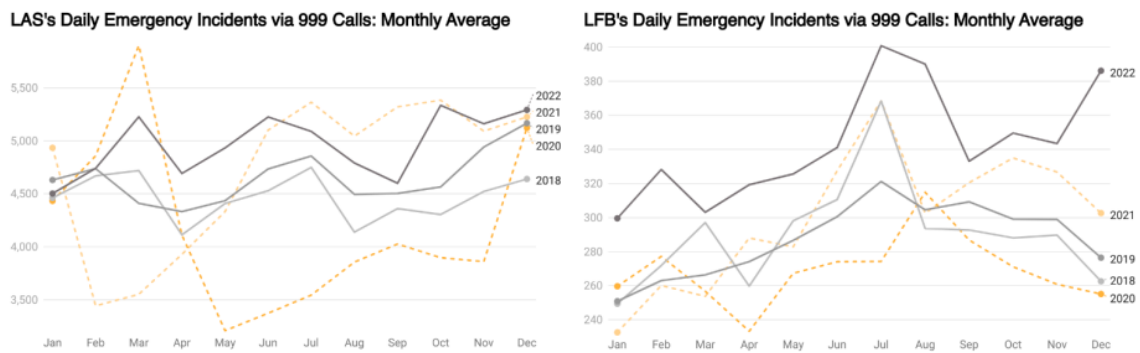

The monthly demand patterns for LFB had remained consistent over the past five years. There was a general upward trend in call volumes during the first half of a typical year until the peak in July, followed by a decline in the second half. However, there were notable increases in demands in October 2021 and December 2022. Furthermore, the demands for both LAS and LFB in 2022 were generally higher than in previous years, indicating an overall increase trend for Blue Light Services.

**Figure 7**
**Average LAS and LFB Demand by Day of Week (Jan 2018–Mar 2023)**

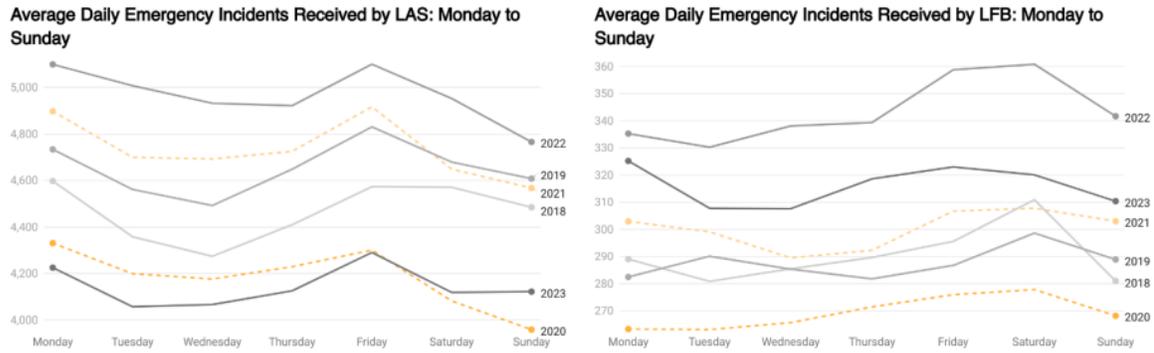

In Figure 7, variations in the average daily demand for LAS and LFB could be observed by day-of-the-week throughout the observation period. LAS call demands (left) exhibited consistent patterns, with higher demands observed on Mondays and Fridays and lower demands on Wednesdays and weekends (Saturdays and Sundays). Similarly, the LFB demands demonstrated similar trends for each year, with relatively lower levels on weekdays but notable increases on Fridays and Saturdays, followed by a decline on Sundays. To further drill down into finer temporal scale, time-of-the-day had been investigated. Figure 8 illustrated the average demand distribution for LAS and LFB by hour from 2018 to 2023. Regions colored in deeper blue represented for working hours, while the lighter blue regions were taken as resting time. Notably, the demands for both services were elevated during working hours (9:00-18:00) and the following four hours until bedtime (18:00-22:00); especially with the peak demand periods between 18:00 and 20:00, which was normally the cooking and dinner hours. It was posited that during such period, individuals often depart from workplaces or residences for social engagements, outdoor activities, or dining at restaurants; hence the increased social activities might relate to higher likelihood of encountering emergencies or requiring urgent medical or fire services. Between 22:00 and 11:00, the demands for LAS and LFB notably subsided below the mean; then from 2:00 to 7:00, the demands for both LAS and LFB reached the lowest levels, in the assumption that most individuals were in deep slumber, hence with fewer societal activities related incidences of emergencies or demands for assistance.

**Figure 8**
**Average LAS and LFB demand by time of day (Jan 2018–Mar 2023)**

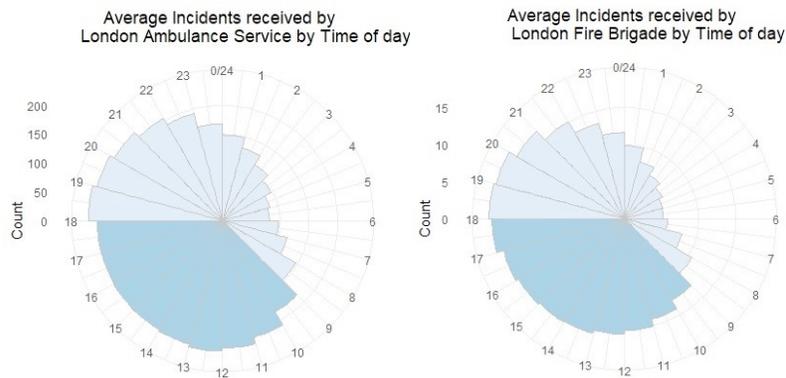

Observations in Figure 5 disclosed potential noisy data and the lacks of distinct visual patterns. To further investigate the long-term seasonal trends of both demands, STL method was employed to decompose monthly time-series data. The Seasonal and Trend decomposition using Loess (STL) applies a repeated loess fitting technique to decompose monthly time-series data into three distinctive components: a smoothed trend, a seasonal component, and a residual component. This decomposition process involves utilizing a continuous loess line to capture the smoothed, long-term trend, alongside employing 12-month-specific loess lines to capture the seasonal variations. The fitting procedure iteratively adjusts these components until the resulting trend and seasonal components converge and exhibit minimal differences from the previous iterations, ensuring an effective time series decomposition into its constituent parts.

**Figure 9**
**STL Decomposition of LAS and LFB Time Series (Jan 2011–Mar 2023)**

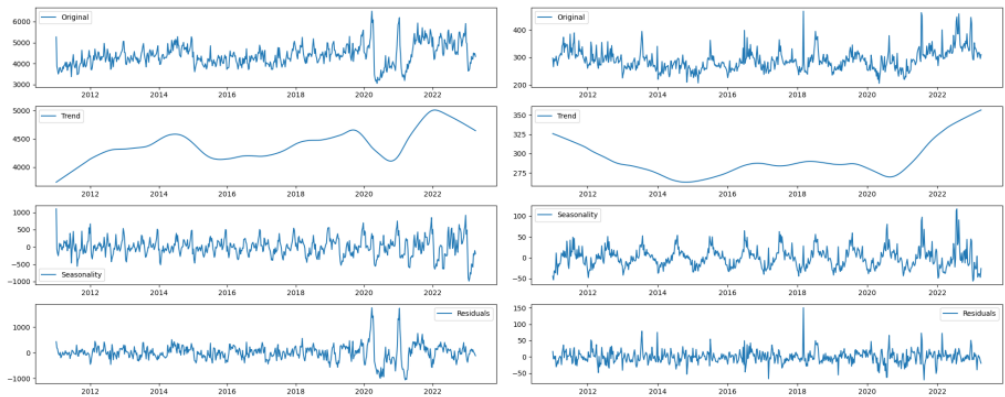

As depicted in Figure 9, the LAS and LFB demands exhibited pronounced seasonal trends, highlighting their cyclic patterns of variations across distinct time intervals. LAS generally peaked during the summer and end of the year, while the beginning of the year tended to mark the nadir of demand. Similarly, LFB received the highest calls in summer, significantly surpassing other timeframes. Such cyclic pattern of seasonal fluctuations spanned approximately a year. Besides, the LAS demand demonstrated a discernible overall upward trend, whereas the LFB demand showcased moments of descending and intermittent ascending trends. Whilst most their residuals exhibited a predominantly stochastic nature.

### 4.1.2. The impact of weather factors on demands of LAS and LFB

Previous studies have consistently demonstrated the significant influence of weather conditions on emergency service demands. Building upon this evidence, our study further investigates how variations in weather conditions relate to the demand patterns of the LAS and LFB, with the aim of identifying shared sensitivities in weather-driven service pressures. An OLS model has been used as the baseline, with temperature, dew point (for humidity), and wind speed as covariates, for respective models' development for LFB and LAS. Based on measures of residuals such as autocorrelation function (ACF) and partial autocorrelation function (PACF), non-stationarity hence temporal dependence had been observed in the residual series. So that it is necessary to introduce the Autoregressive Integrated Moving Average with Exogenous Factors (ARIMAX) model, which is capable of addressing temporal autocorrelation; and the Seasonal ARIMAX (SARIMAX) model incorporating seasonality. By employing a grid search technique, model with the best fitting quality has been selected by the smaller AIC value.

**Table 1**
**Evaluation Metrics for LFB and LAS Models**

| Data | Model | Adj R-square | AIC | RMSE |
|------|-------|--------------|-----|------|
| LFB | OLS | 0.24 | 6314 | 33.37 |
| | ARIMAX | 0.51 | 5908 | 26.77 |
| | SARIMAX | 0.44 | 5515 | 28.71 |
| LAS | OLS | 0.01 | 9803 | 509.48 |
| | ARIMAX | 0.57 | 8976 | 335.72 |
| | SARIMAX | 0.5 | 8364 | 363.9 |

In Table 1, time series models outperformed the OLS model, with the adjusted R-squared for LFB nearly doubling. ARIMAX and SARIMAX models improved the adjusted R-squared for LFB from 0.24 to 0.51 and 0.44 respectively, whilst LAS saw an increase from 0.01 to 0.57 and 0.50 for its R-squared correspondingly. The root mean squared difference (RMSE) between the model's predicted values and the actual observations, provided insight into the model's adherence to the original data. RMSE values decreased in both the ARIMAX and SARIMAX models, with a particularly significant reduction observed in the LAS model, which decreased by nearly half. AIC measure has been utilised to reflect the relative amount of information the model compromises by considering the trade-off between overfitting and underfitting. In the case of both the ARIMAX and SARIMAX models, a substantial decrease in AIC was observed indicating the retention of a substantial amount of relevant information, while its residuals from either model did not exhibit significant autocorrelations at various lags, indicating that the current models have effectively captured the primary temporal features.

**Table 2**
**Coefficient estimates for LAS Models vs. LFB Models**

| ARIMAX | | | SARIMAX | | |
|---|---|---|---|---|---|
| Parameter | LAS Estimate | LFB Estimate | Parameter | LAS Estimate | LFB Estimate |
| Temperature | 53.25*** | 6.13*** | Temperature | 63.52*** | 6.53*** |
| Dew Point | -10.47 | 4.35*** | Dew Point | -14.76(.) | 5.43*** |
| Wind Speed | -5.83(.) | 1.08*** | Wind Speed | -7.02** | 1.09*** |
| ar.L1 | 0.86*** | 0.35*** | ar.L1 | -0.99*** | 0.26*** |
| ma.L1 | -0.93*** | 0.91*** | ma.L1 | 1*** | 0.92*** |
| ma.L2 | -0.06 | | ma.S.L52 | -0.77*** | 0.84*** |

*Signif. codes: 0'***'  0.001'**'  0.01'*'  0.05'.'  0.1''*

Further summarised in Table 2, both the ARIMAX and SARIMAX models for LAS and LFB demands included an autoregressive component (AR1) and a moving average component (MA1), representing for a close correlation between current demand and prior demand as well as potential error terms; in addition, SARIMAX model also featured a significant seasonal moving average term (SMA 52), indicating that LAS demand exhibited notable annual cyclical error correlations, with demand patterns significantly influenced by yearly seasonal factors. Weather factors hereby had been considered as exogenous variables for both LAS and LFB demands. (1) Temperature, in particular, had a significantly positive correlation with demand for both services, while dew point and wind speed displayed negative correlations. Besides, the impact from weather on LAS demand was more pronounced, especially for temperature and dew point. (2) Temperature was observed as a significant impact for LFB demand as well, surpassing effects from dew point and wind speed. Specifically, with each unit increase in temperature, LFB demand was projected to increase by 6.13 units. In such climatic contexts, vegetation and combustible materials become susceptible to moisture loss, particularly plants and trees that can serve as potential fuel sources. Gill's study in Australia's World Heritage-listed Kakadu National Park similarly identified an elevated fire risk under low dew point conditions [50]. Moreover, it is plausible that higher temperatures could trigger fires through power lines or mechanical equipment. In hot and arid conditions, the spread of fires is also more likely to escalate rapidly [51, 52]. It is evident that the absolute estimate of the dew point escalated to 5.43 in the SARIMAX model, suggesting a heightened negative correlation of dew point, hence LFB demand was more pronounced under drier climatic conditions. (3) the parameter estimates for temperature in the ARIMAX and SARIMAX models for LAS demand, were remarkably higher than those for dew point and wind speed. The absolute values of the coefficients related to weather factors had increased in the SARIMAX analysis in that, the temperature coefficient had risen from 53.3 to 63.5, while the dew point coefficient had shifted from -10.47 to -14.76. This indicated that a one-unit increase in temperature corresponded to a 63.5-unit increase in LAS demand, while a one-unit decrease in dew point led to a 14.76-unit increase in LAS demand.

In all, elevated temperatures can moderately affect human health, potentially leading to heat-related ailments such as heat exhaustion and heatstroke, thereby increasing the frequency of LAS calls. Furthermore, this pattern could be attributed to substantial variations in dew point across different seasons. A lower dew point signifies drier air, facilitating the dissemination of viruses and bacteria as minute particles. The accelerated moisture evaporation in low humidity conditions renders airborne microorganisms more prone to existence and transmission, thus extending their propagation distance and duration. In a similar context, Brzezińska-Pawłowska's research also emphasized the impact of seasonal factors on ambulance demand. Furthermore, their study revealed a correlation between dew point and exacerbations of asthma and COPD conditions [53]. On the other hand, wind speed exhibited a negative correlation with LFB demand, which contrasts with the typical expectation that higher wind speeds exacerbate fire spread. This counterintuitive result may reflect increased vigilance and precautionary measures during strong winds, reducing fire incidents. Furthermore, the significant proportion of LFB calls related to special services might weaken the direct link between wind speed and fire risk. For LAS demand, the negative correlation was likely due to high wind speeds discouraging outdoor activities, thereby reducing the likelihood of accidents.

## 4.2. Spatial patterns analysis of LAS and LFB

### 4.2.1. Bivariate maps: All years, COVID v.s. Non-COVID periods

The bivariate map in Figure 10 illustrated LAS and LFB demands by utilizing colour codes to indicate demand intensity, where darker shades denote higher demand, while lighter tones represent lower demand. Table 4 presents the top ten administrative boroughs with the highest demand from the LAS and the LFB. The rankings are based on three standardised indicators: total LAS call volume, total LFB call volume, and the number of LSOAs with high demand for both LAS and LFB, weighted at 0.4, 0.4, and 0.2 respectively. Total call volume is given greater weight as it directly reflects overall workload. Figure 10 illustrates the areas where LAS and LFB demands overlap; as shown, the dark purple areas—indicating high demand for both services—radiate outward from London's central boroughs. Among the top ten boroughs listed in Table 3, eight are located in central London, with only Croydon and Hillingdon situated in Outer London.

<table>
<tr><td align="center"><strong>Figure 10</strong><br><strong>Bivariate map of LAS and LFB demands (Jan 2018–Mar 2023)</strong></td><td align="center"><strong>Table 3</strong><br><strong>Top 10 Boroughs for LAS and LFB<br>(Jan 2018–Mar 2023)</strong></td></tr>
</table>

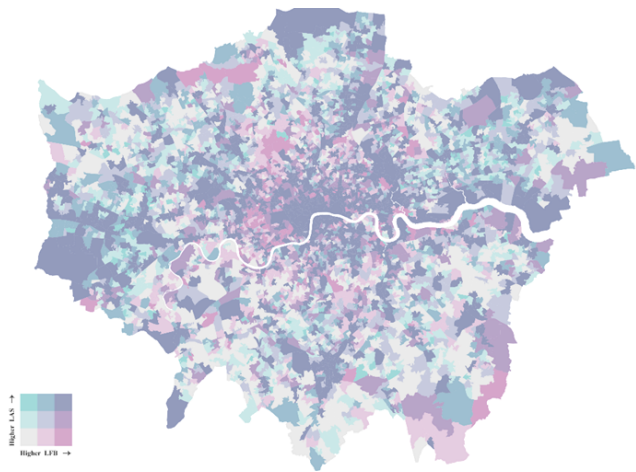

| Top 10 Borough |
| --- |
| Westminster |
| Tower Hamlets |
| Southwark |
| Camden |
| Croydon |
| Lambeth |
| Newham |
| Islington |
| Hillingdon |
| Hackney |

It was also interesting to find outskirt areas near Heathrow Airport, the northern extents of Enfield, and both the northern and southern areas in Havering displayed heightened demands for both services. Moreover, centrally located regions often exhibited higher LFB and lower LAS demand; with exceptions such as Bromley's peripheral zones and Barnet's central areas, indicating increased LFB demand and lower LAS demands. In peripheral regions, LAS demand typically surpassed LFB demand, or both were relatively modest. From time series analysis of LAS and LFB, significant fluctuations in LAS demand were observed during the COVID-19 pandemic, making comparative analysis being crucial between periods during the pandemic and non-pandemic (pre- or post-). In Figure 11, bivariate maps were generated to visually depict these comparisons, with the left map illustrating the pandemic period, and the right map representing the non-pandemic periods; followed by Table 4 outlined the top 10 borough areas with higher LAS and LFB demands during the respective periods. It was obvious that certain regions were outlined with black borders on the left, indicating for their non-high demand for LAS and LFB services during the non-pandemic period. However, during the pandemic, they transformed into regions with elevated demand for both services. Conversely, the right map highlighted specific areas similarly outlined in black, which were characterized by simultaneous high demand for both LAS and LFB services during the non-pandemic period but without such uniformity during the pandemic.

**Figure 11**
**Bivariate maps of LAS and LFB demands: COVID vs. Non-COVID Periods**

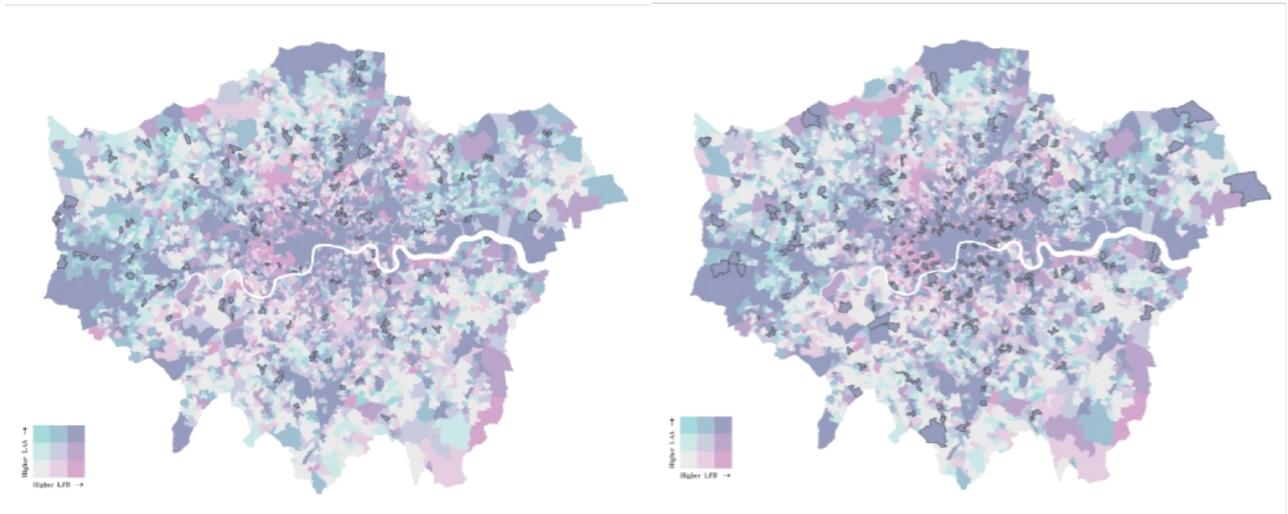

**Table 4**
**Top 10 Boroughs for LAS and LFB: COVID vs. Non-COVID Periods**

| Top 10 Borough | |
| --- | --- |
| **Covid** | **Non-Covid** |
| Westminster | Westminster |
| Southwark | Southwark |
| Tower Hamlets | Tower Hamlets |
| Croydon | Camden |
| Camden | Croydon |
| Newham | Lambeth |
| Haringey | Newham |
| Hackney | Islington |
| Islington | Hillingdon |
| Hillingdon | Hackney |

It seems that that demand distributions mostly remained consistent with the general patterns observed in Figure 10. High-demand areas for services were concentrated in central London, with fewer instances in peripheral areas. However, the demands for LAS and LFB services were higher in central areas during non-pandemic periods than the pandemic period, with newly outlined areas emerging in central London. Figure 11 also highlighted a shift from certain LSOAs during the pandemic for higher demand for LAS and LFB services: LSOAs with low LAS demand but high LFB demand in central areas, and LSOAs with high LAS demand but low LFB demand in peripheral regions. It was apparent that from Table 5, boroughs such as Croydon, Newham, and Hackney saw an increase of demands in their rankings during the Pandemic period, while Camden, Lambeth, and Islington ranked higher in demands during the non-Covid period.

**Figure 12**
**Monthly Trends of LAS and LFB Demand in High-Demand LSOAs during Non-COVID Periods**

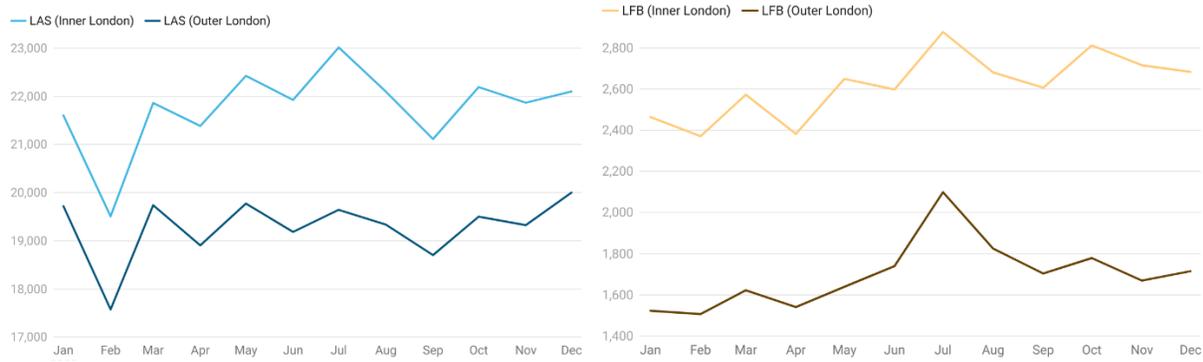

Given that service demand patterns exhibited abnormal fluctuations during the pandemic, the spatial distribution during non-pandemic periods was used to better represent the regular demand for emergency services. This study further identified LSOA areas with high demand for both LAS and LFB during non-pandemic periods and analysed their monthly trends.

The results indicated that demand in Inner London was consistently higher than in Outer London. In the high-demand areas of Inner London, both fire and ambulance services experienced significant peaks in July, with secondary peaks observed in March, May, and October—though these were less significant than in July. In contrast, selected areas in Outer London showed a clear LFB demand peak only in July, while LAS demand exhibited multiple peak points not only in July but also in January, March, May, and December.

**Figure 13**
**Monthly Trends of LAS and LFB Demand in High-Demand LSOAs during COVID Periods**

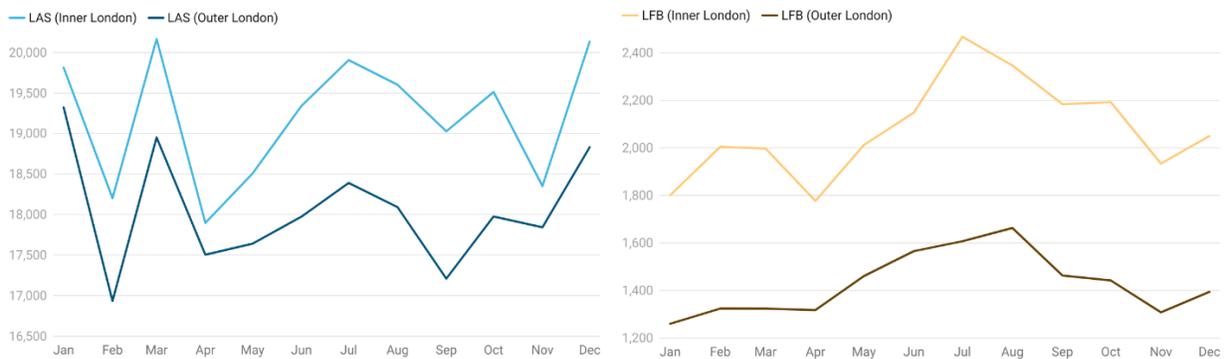

During the pandemic, seasonal peaks for LFB in the high-demand areas remained concentrated in the summer months, while LAS peaks shifted to January, March, and December. Notably, Inner London still showed an upward trend in LAS demand in July, whereas this trend was less evident in Outer London. Overall, LAS's seasonal pattern was significantly disrupted during the pandemic, whereas LFB's remained consistent.

**Figure 14**
**LISA Map of LAS and LFB Demands (Jan 2018–Mar 2023)**

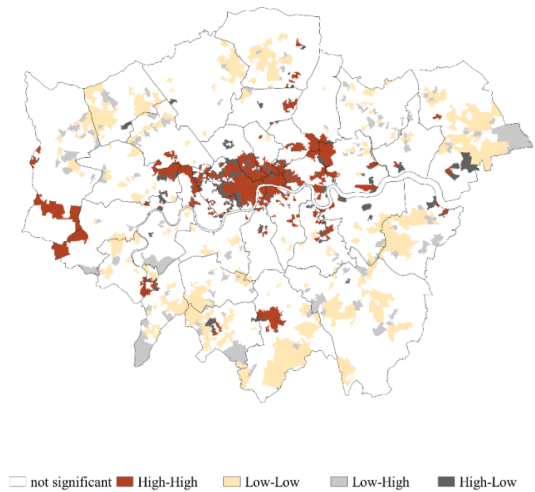

not significant ☐   High-High ■   Low-Low ☐   Low-High ☐   High-Low ■

Upon assessing Bivariate Moran's *I* for LAS and LFB demands, LISA maps could be generated (Figure 12) to further validate their significant spatial dependence. This correlation remained consistent across different temporal contexts from 2018 to 2023 (value at 0.24), regardless of the period whether being in the pandemic (value at 0.181), or the non-pandemic periods (value at 0.252, while all Moran's *I* values > 0, p-value < 0.01). Such finding suggested that LAS and LFB exhibit specific analogous spatial patterns: when LAS demand was elevated, the surrounding areas also tended to experience heightened LFB demand, and vice versa. Notably, during the pandemic period, the Bivariate Moran's *I* value for LAS and LFB demand decreased to 0.177, which was lower than the non-pandemic periods. This observation might indicate that the pandemic induced fluctuations concurrently moderated their interrelatedness. Figure 12 illustrated the spatial correlations between LAS and LFB demands from January 2018 to March 2023. Hot spot (High-High) regions primarily concentrated in the city centre, particularly in Westminster, the City of London, and Camden, but also found clusters around the periphery of Heathrow Airport and in the western-central portion of Croydon. Cold spot (Low-Low) and outliers (Low-High) regions were more widely distributed and tend to co-occur, mainly in peripheral areas, with a notable concentration in the southern regions, such as in Croydon, Bromley, Bexley, and Havering.

**Figure 15**
**LISA Maps of LAS and LFB Demands: COVID (left) vs. Non-COVID (right) Periods**

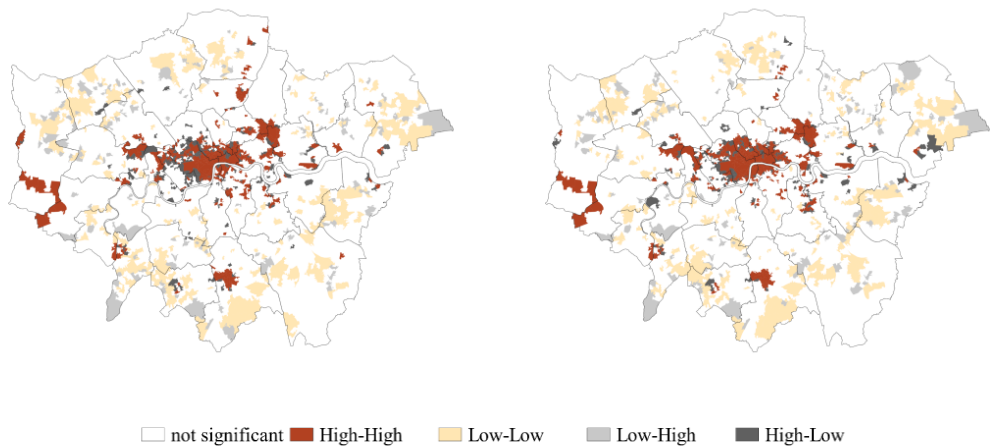

not significant ☐   High-High ■   Low-Low ☐   Low-High ☐   High-Low ■

To further evaluate the influence on demands from the Pandemic, comparative LISA maps had been produced in Figure 13. It became clear that the occurrence of hot spot clusters in the central region decreased during the pandemic, while the northern areas of Haringey and Enfield saw an increased prevalence of hot spot clusters during the pandemic. Additionally, there was a rise in cold spot clusters, particularly in the northwest, covering regions such as Barnet, Harrow, and Hillingdon. By now, we could confirm the existence of a positive spatial correlation in demand distribution between LAS and LFB emergency services in London, coupled with the manifestation of significant clustering phenomena in specific regions. The overarching distribution trends across the three periods were similar, with high-demand hot spot regions of LAS and LFB predominantly converging in the central heart of London and sporadically near Heathrow Airport. In contrast, peripheral areas primarily comprised cold spot low-demand regions or areas characterized by elevated LAS demand and subdued LFB demand.

### 4.2.2. Bivariate Maps: LAS and LFB Incidents by Type

The LFB provides a range of services, such as responses to flooding, elevator rescues, road traffic collisions, and false alarms with fire incidents constituting approximately 15%. The three bivariate maps in Figure 14, respectively depicted demand of LAS and LFB for fire incidents, special services, and false alarms. In general, areas with high demand for fire incidents and LAS services covered a broader range than those of the other two. Regions with high call volumes for fire incidents and LAS services were mainly focused in the city centre, alongside the eastern bank of the Thames River, and in the Hillingdon area, particularly near Heathrow Airport. Further rankings presented in Table 5 highlighted elevated demand central boroughs and Ealing, Brent, Croydon, and Haringey. It was also notable that the outskirts of London, particularly in the southern part of Bromley, the northwestern region of Hillingdon, as well as Barnet and Bexley, showed higher demand for fire incidents but lower demand for LAS services.

**Figure 16**
**Bivariate maps of LAS and LFB demand types – Fire (Left), Special service (Middle), False alarm(Right)**

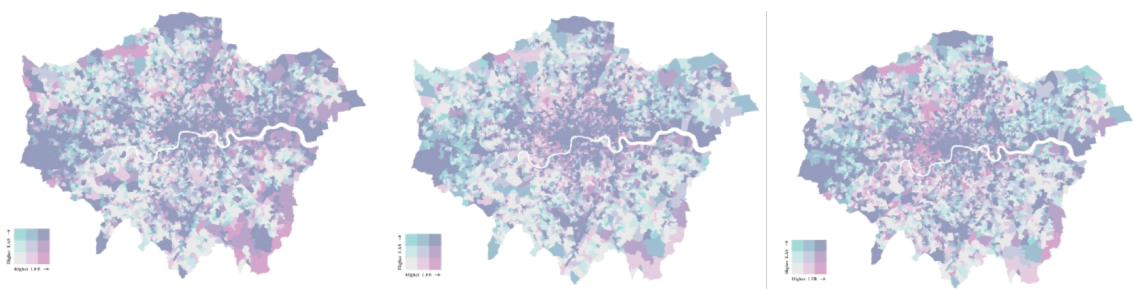

**Table 5**
**Top 10 boroughs for LAS demand and different types of LFB demand**

| Top 10 Borough | | |
| --- | --- | --- |
| LFB (Fire) and LAS | LFB (Special Service) and LAS | LFB (False Alarm) and LAS |
| Tower Hamlets | Tower Hamlets | Westminster |
| Southwark | Southwark | Southwark |
| Newham | Westminster | Tower Hamlets |
| Westminster | Islington | Camden |
| Hillingdon | Croydon | Croydon |
| Ealing | Lambeth | Lambeth |
| Brent | Hackney | Islington |
| Croydon | Camden | Ealing |
| Haringey | Haringey | Newham |
| Hackney | Hillingdon | Hillingdon |

High-demand areas for special LFB services and LAS exhibited a distinctive radiating pattern, with dense concentrations in the centre and a sparser distribution toward the periphery. As indicated in Table 5, central boroughs such as Tower Hamlets, Southwark, and Westminster consistently ranked highest in demand, with Islington, Lambeth, and Camden also ranking highly, alongside peripheral hotspots like Heathrow and Croydon. It could be also observed that the distribution of LAS calls and false fire alarms closely aligned with the all-years distribution seen in Figure 5 for all calls, potentially because false fire alarms constituting approximately half of the LFB incidents. Additionally, some central-west areas were more prone to a scenario where false fire alarms are abundant and LAS demand was lower, such as in Westminster, Kensington and Chelsea, and Camden.

Throughout the observation period, LAS's spatial distribution remained relatively stable, whereas the spatial distribution of different LFB service types varied. These differences resulted in spatial inconsistencies among LFB high-demand areas that overlapped with LAS, across different service types. Nevertheless, approximately 61% of LSOAs identified as dual high-demand areas overlapped, suggesting a strong spatial coupling between LFB and LAS high-demand zones despite differences in LFB service types.

**Figure 17**
**Spatial Overlap of High-Demand Areas Between LAS and LFB Service Types Across London**
*Fire (Left),  Special service (Middle),  False alarm (Right)*

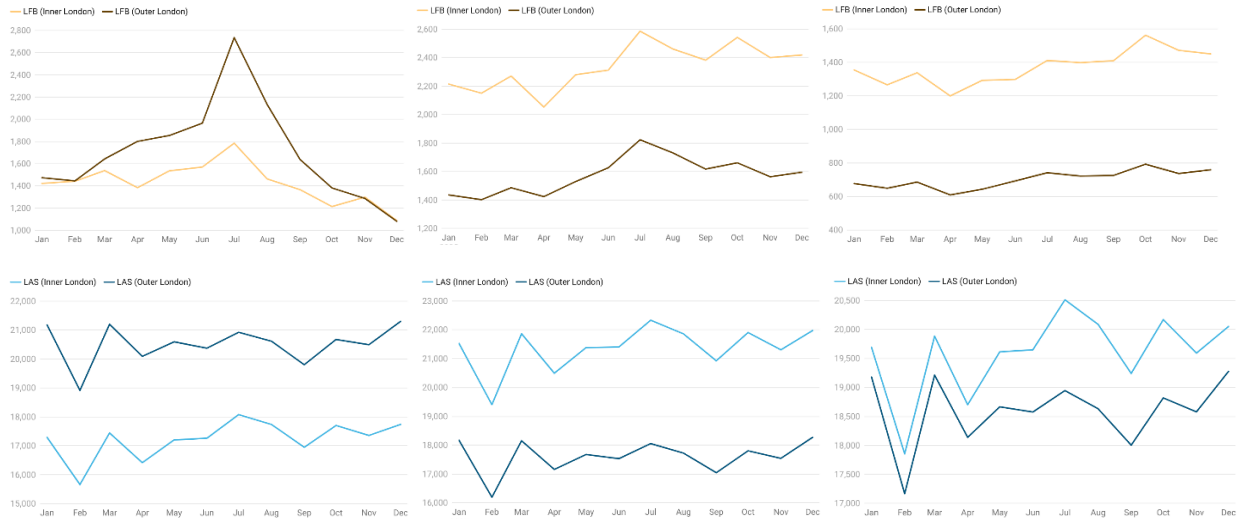

As detailed in Section 4.2.1, the study further categorised LSOAs into three groups based on different LFB service types that matched high LAS demand and compared their monthly trends. In areas where fire incidents and LAS demand were both high, Outer London exhibited a greater overall service demand than Inner London. Both services peaked in July, with LAS demand showing a more marked increase in Inner London, while the rise in Outer London was comparatively modest. In special LFB service and LAS high-demand areas, Inner London showed peaks in July and October, whereas Outer London demonstrated significant alignment only in July. In areas where False Alarms and LAS were both in high demand, monthly trends for false alarms were relatively consistent across the city, with a peak in October and a smaller rise in July. LAS also showed a mild peak in October, while Inner London exhibited a more notable increase in July. Despite differences in overall trends, the two services exhibited partial alignment at certain time points.

Follow-up bivariate Moran's *I* analyses, a significant positive spatial autocorrelation among all three types of LFB demand and LAS demand. It indicated that fire incidents (value at 0.227), specialized services (value at 0.205), false alarms (value at 0.223), and LAS demand were spatially correlated, with similar values clustering together rather than being randomly distributed. Broadly, higher LAS demand tends to coincide with higher LFB demand in the surrounding areas, and vice versa. It was also noteworthy that the Moran's *I* value for fire incidents in LFB demand and LAS demand was the highest, followed closely by false alarms, then specialized services, suggesting that the correlation between the distribution of fire incidents and LAS demand was the strongest.

Comparative LISA maps in Figure 15 were generated to further assess differences among LFB services. It became evident that hot spot regions between LAS demand and fire incident demand was primarily concentrated in the city centre and the western region along the Thames River, with additional clustering in southern Hillingdon, HH clustering for special service and false alarm incidents with LAS demand were mostly centred in the urban core, with only a few clusters near Heathrow Airport and western Croydon. It is worth mentioning that hot spot clusters of special service demand was more broadly distributed within the central region and includes 421 LSOAs, compared to 357 for fire incidents and 232 for false alarms, which had the fewest hot spot clusters.

**Figure 18**
**LISA Maps: LAS and LFB demand types – Fire(Left), Special service(Middle), False alarm(Right)**

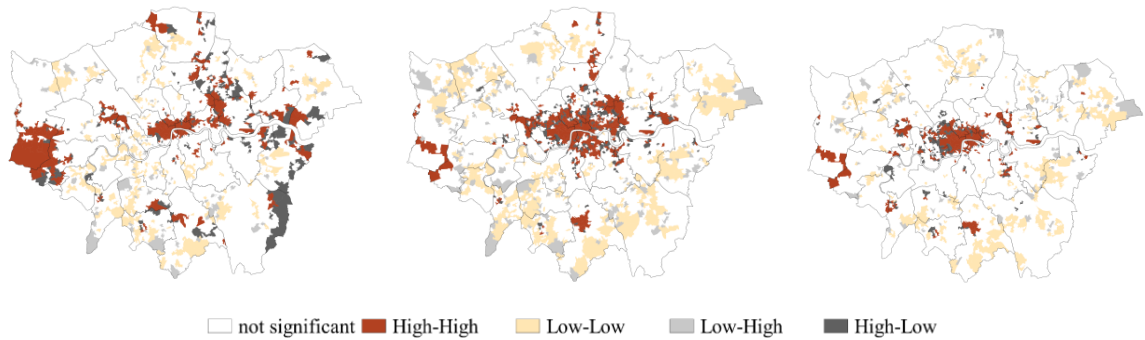

not significant ■ High-High ▨ Low-Low ▨ Low-High ■ High-Low

The cold spot cluste... ...e et al.) were mainly concentrated on London's outskirts, with special service inci... ...in the southern regions, such as Havering and Harrow, while false alarms and fir... ...ore dispersed. It can be observed that high-value outliers for fire incidents are con... ...nley and alongside the Thames River, while fewer high-value outliers for special ... ...r hot spot clusters in the city centre. It also became clear that low-value outliers for... ...distributed in the eastern and southern peripheries, whereas LH clusters for False Ala... ...ns in the northwest and east.

Pr ... ...characteristics and emergency service demand... ...ing the spatial and temporal patterns of emergency incidents. Building on this body of evidence, the present study aims to identify and compare the socio-economic determinants influencing the demand for LAS and LFB services, with particular attention to the similarities and differences in their driving factors.

**Table 6**
**Evaluation metrics: OLS vs. GWR for LFB and LAS**

| Data | Model | Adj R-square | AIC |
|------|-------|--------------|-----|
| LFB | OLS | 0.59 | 6192 |
| | GWR | 0.69 | 5623 |
| LAS | OLS | 0.24 | 12424 |
| | GWR | 0.48 | 11148 |

Standardized variables had been incorporated into OLS regression model and GWR regression model, to examine the influence of model performance from presence of spatial autocorrelation. The results consistently indicated significant spatial dependence in the residuals of both models (p-value < 0.01), and suggested that the OLS model fails to capture spatial variability and heterogeneity across geographical locations. Therefore, the GWR model, by accounting for spatial heterogeneity, offers a more accurate representation of spatial relationships and better handles spatial dependence than the OLS model.

Table 6 presents the results from the OLS and GWR models, where the GWR models for both LAS and LFB significantly outperform the global regression models. The adjusted R-squared, which signifies the proportion of variance explained by the model, emphasises that LAS's GWR model can account for 48% of the variance, and LFB's GWR model can explain 69% of the variance. In addition, the GWR models hold substantial reductions in AIC values, suggesting a better pertaining with crucial information hence superior model fit.

*4.2.3.1. GWR modelling on LAS demand patterns*

In LAS's GWR model, most selected variables, except for "Road Casualty", had significant effects along the Thames River coastline, in areas such as Greenwich, Newham, and Tower Hamlets; in the northwest regions, like Barnet and Harrow, factors such as the proportion of residents without qualifications, Fair Health ratios, and the IMD Health Deprivation score played significant roles onto LAS demands. Factors posed positive influences onto the LAS demand's increases were the proportion of residents without education, PTAL, and "Road Casualty". Especially the effect from "Road Casualty" was spreading throughout the city centre, southwest regions like Harrow, and extended to areas such as Kingston upon Thames and Croydon. For example, a one-standard-deviation increase in road accidents corresponds to an increase in LAS demand by 1.7 to 2.7 times (0.2 to 0.6 in log-transformed units). In eastern areas along the Thames (e.g., Greenwich, Newham, and Tower Hamlets), the proportion of residents without education and PTAL positively influenced LAS demand, while in western regions like Richmond and Hounslow, PTAL showed the strongest effects.

The relationships between LAS demand and other factors then vary by location. For example, the proportion of elderly residents might correlate positively with LAS demand, but turned to be negatively influential in areas along the Thames River (except for Newham). Similarly, the proportion of residents in "Fair" health may increase demand of LAS in Haringey, Richmond upon Thames, and Wandsworth, but reduced the LAS demands in Barnet and Tower Hamlets. For the same borough, proportion of households without cars instead could drive LAS demand high, but the same factor reduce such demands in Greenwich, Newham, and Wandsworth. A similar pattern was witnessed with IMD living environment scores, which positively correlate with LAS demand near Greenwich and Newham but turned negative along the Thames coastline and in Barnet, with weaker negative effects closer to the river. Lastly, single-person household proportions and IMD Health Scores exhibited positive correlations with LAS demand, especially in Greenwich, Newham, and areas near the Thames like Wandsworth and Hammersmith & Fulham. However, negative correlations for single-person households were found in parts of Newham and Lambeth, while the IMD Health Score showed negative correlations mainly around Greenwich.

**Figure 19**
**Distribution of coefficient estimates in LAS GWR model**

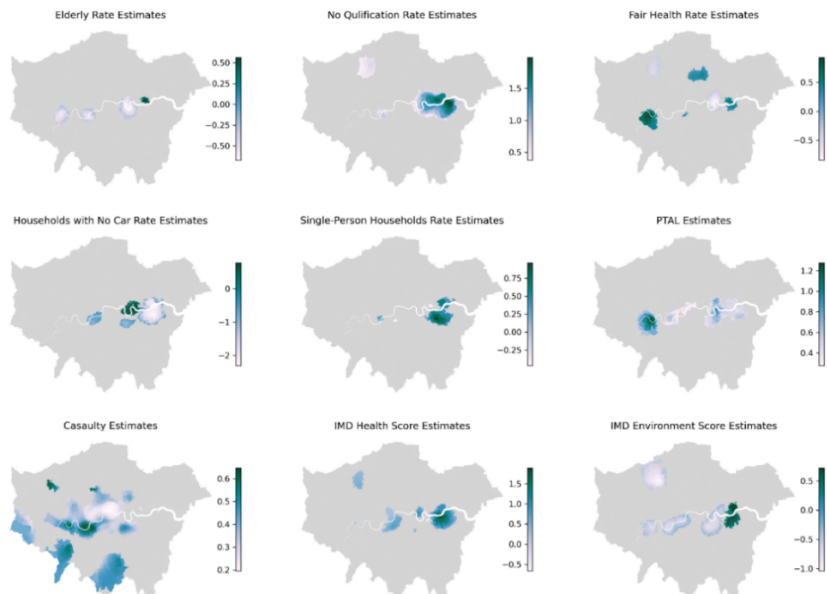

**Table 6**
**Coefficient estimates of LAS GWR model**

| Dependent Variable | Independent Variable | Mean | STD | Min | Median | Max |
|---|---|---|---|---|---|---|
| | Intercept | 0.086 | 0.398 | -1.867 | 0.118 | 1.085 |
| | Elderly (%) | -0.021 | 0.117 | -0.672 | 0.002 | 0.565 |
| LAS | No Qualification (%) | 0.197 | 0.295 | -0.575 | 0.144 | 1.982 |
| | Fair Health (%) | 0.091 | 0.163 | -0.847 | 0.091 | 0.933 |
| | Households with No Car (%) | 0.004 | 0.369 | -2.306 | 0.068 | 0.779 |
| | Single-Person Households (%) | 0.126 | 0.135 | -0.466 | 0.125 | 0.965 |

| | Mean | STD | Min | Median | Max |
|---|---|---|---|---|---|
| PTAL | 0.162 | 0.164 | -0.256 | 0.128 | 1.277 |
| Casualty | 0.267 | 0.133 | -0.342 | 0.261 | 0.646 |
| IMD Health Score | 0.168 | 0.22 | -0.662 | 0.134 | 1.884 |
| IMD Environment Score | -0.08 | 0.215 | -1.039 | -0.034 | 0.717 |

The GWR model reveals several key relationships between covariates and LAS demand. As expected, road casualties showed a positive correlation with LAS demand, as traffic accidents typically require urgent medical care. Similarly, areas with lower educational attainment were positively associated with emergency calls, consistent with Proctor's findings [57]. Lower education levels may lead to medical uncertainty hence increase the likelihood of emergency calls. On contrary to expectations, PTAL also correlated positively with LAS demand, in line with Burns' study on that, individuals who frequently utilize emergency services prefer ambulance transportation over complex public transportation for emergencies [58]. The proportion of single-person households was assume to be related to frequent emergency calls due to the absence of immediate assistance in emergencies, which is consistent with Siler's findings [59]. However, regions exhibiting a negative correlation may be influenced by more joint factors. Such mixed influences were also found in, for example the positive correlation between the proportion of elderly residents and LAS demand, but with a negative relation along the Thames River. Similarly, the proportion of residents in "Fair" health correlated positively in areas like Haringey, Richmond upon Thames, and Wandsworth but being negative in Barnet and Tower Hamlets. Furthermore, the relationship between the proportion of individuals without access to a car and the demand for LAS showed varying positive and negative correlations, with areas receiving negative influences might in favor of alternative transportation modes.

### 4.2.3.2. Interpretation of the GWR model for LFB demand patterns

Figure 17 and Table 7 showed that all selected variables, except for "Home Ownership Rate", might increase demands for LFB. The proportion of single-person households showed a strong correlation, especially in Outer London areas, such as Bexley, Barnet, and Harrow. Road incidents could also drive LFB demand high, especially in southeastern regions like Bromley and Croydon, where one standard deviation increase was expected to boost LFB demand by 1.6 to 2.1 times. PTAL was positively correlated with LFB demand, but its statistical significance was only observed in southwest Hounslow and along the Thames in Westminster. Additionally, IMD crime scores could increase LFB demands in the eastern and northern regions, but it reduced such demands in parts of Westminster. The only variable negatively correlated with LFB demand was the proportion of households owning properties. One standard deviation increase would result in a decrease in LFB demands by 1.8 to 2.2 times in areas near the river in Hounslow and Ealing.

The positive correlation between the proportion of single-person households and LFB demands could be attributed to higher levels of social isolation and safety concerns in areas with a significant number of single-person households, leading to more frequent reliance on fire brigade assistance during fires or other emergencies. This observation contrasted with Corcoran's findings in South Wales, which indicated that densely populated households were more likely to trigger fire calls [49], but aligned with Corcoran's research in Brisbane and Cardiff, suggesting that areas with lower property ownership rates tend to exhibit higher fire service demands [57]. Additionally, PTAL could increase LFB demands by facilitating quicker response times in well-connected areas, prompting residents to seek more fire brigade support. Furthermore, the positive influence of road incidents on LFB demands could be explained by the LFB's responsibilities in addressing road traffic accidents. The negative correlation between IMD crime scores and LFB demands suggested that areas with higher crime risks may have had an unstable social environment and strained community relations, potentially driving an increase in arson incidents. This finding was consistent with Wuschke and Corcoran's research, which linked fire incidents to community deprivation and disadvantage [26, 49].

**Table 7**
**Coefficient estimates of LFB GWR model**

| Dependent Variable | Independent Variable | Mean | STD | Min | Median | Max |
|---|---|---|---|---|---|---|
| | Intercept | 4.531 | 0.192 | 3.131 | 4.524 | 5.153 |
| LFB | Single-Person Households (%) | 0.183 | 0.125 | -0.239 | 0.188 | 0.512 |
| | Home Ownership (%) | -0.256 | 0.15 | -0.825 | -0.252 | 0.228 |
| | PTAL | 0.041 | 0.14 | -0.464 | 0.037 | 0.866 |

| | | | | | |
|---|---|---|---|---|---|
| Casualty | 0.291 | 0.135 | -0.137 | 0.267 | 0.919 |
| IMD Crime Score | 0.122 | 0.117 | -0.252 | 0.116 | 0.545 |

**Figure 20**
**Distribution of coefficient estimates in LFB GWR model**

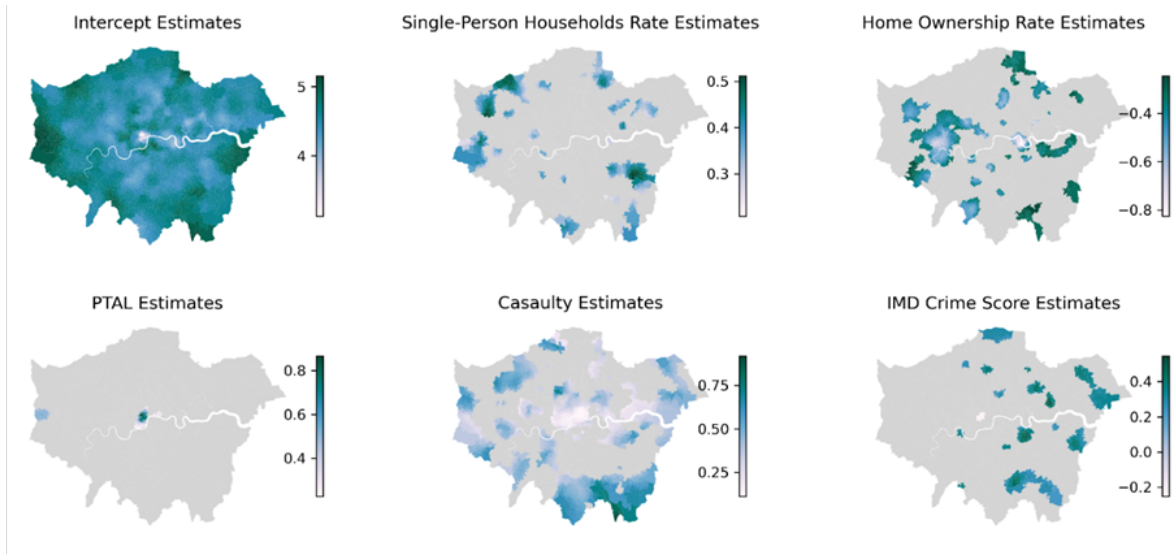

## 4.3. Temporal and spatial distribution of LFB

Temporal distribution of the three types of LFB demands were depicted in Figure 18, presenting consistent distribution pattern throughout the day by hour. Demand remained relatively low in the morning before 9 AM with lowest around 5AM, but increased from 9 AM to 7 PM, then gradually tapered off after 7 PM. Aggregating onto weekly basis, it turned out that weekdays have higher False Alarms than real calls for Fire and Special Services, but with converse pattern for either on weekends. When aggregated to monthly basis, the demands for Special Service and False Alarm incidents peaked in March but were lowest levels in April. During other periods, fluctuations occur within a specific range. The demand for Special Services and Fire-related demands both peaked during the summer, with relatively lower demand in other time frames.

To explore spatiotemporal variations, we employed univariate and bivariate maps at different temporal scales (monthly, daily, and hourly) using comap visualization. The investigation examined distribution of Fire incidents, then categorized into Primary Fire and Secondary Fire. The former incidents presented minimal variation across different quarters and weekdays, but became apparent that (as depicted in Figure 19), Greenwich area along the Thames River experienced higher demand intensity in the afternoon, while demand density decreases in the northern part of the Croydon area during the evening.

**Figure 21**
**Trends in LFB demand by month and hour for incident types**

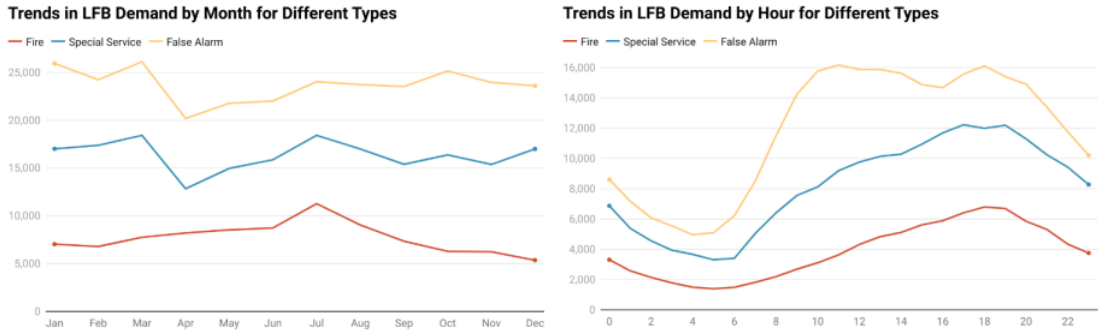

**Figure 22**
**Spatial distribution of primary fires by time of day**

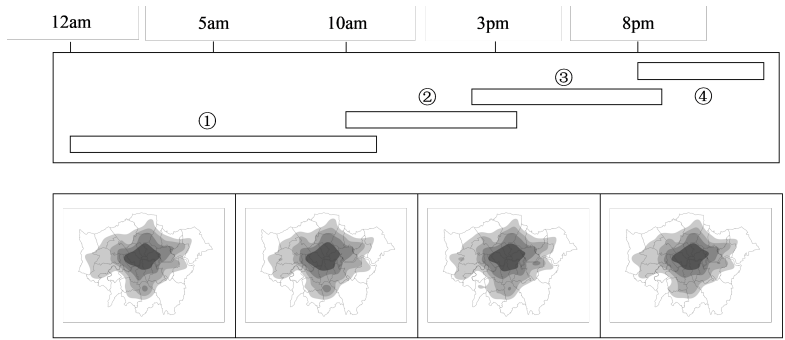

Secondary fire incidents (Figure 20), on the other hand, demonstrated significant spatial variations on monthly and hourly bases. From June to September, there was a broader coverage with larger areas of high demand intensity. In contrast, the intensity range became more confined from October to February. During the summer, heightened demand intensity was concentrated in specific peripheral regions, i.e., the convergence of Ealing and Hounslow, the intersection of Greenwich and Bexley to the east, and the juncture of Croydon and Sutton in the south. Moreover, the distribution between October and February showed relatively consistent intensity levels across different times of the day, while pronounced differences were evident during the other two monthly periods. Remarkably, peak intensities around midday and afternoon extended towards the northeast, forming two interconnected high-density clusters, but with increases in intensity in the western and southern peripheries.

**Figure 23**
**Spatial distribution of secondary fires by time of day and month**

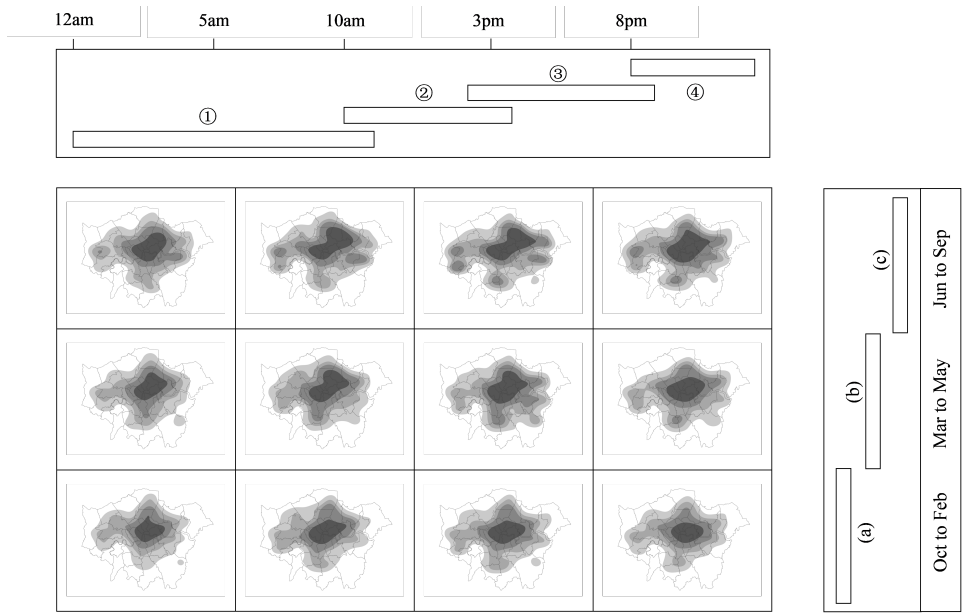

The distribution of special service demands and false alarms showed no evident changes by month or hour (Figure 5,6 and 6 in the Appendix). However, an intriguing finding emerges concerning Flooding incidents within the special service category. In Figure 21, the hourly distribution demonstrated consistent peripheral distribution, while the central high-density area exhibited significant variations. The morning and afternoon had the core region located on the western side of Inner London around Westminster, while by noon, three high-density clusters formed in the northeast.

**Figure 24**
**Spatial distribution of flooding incidents by time of day & by month**

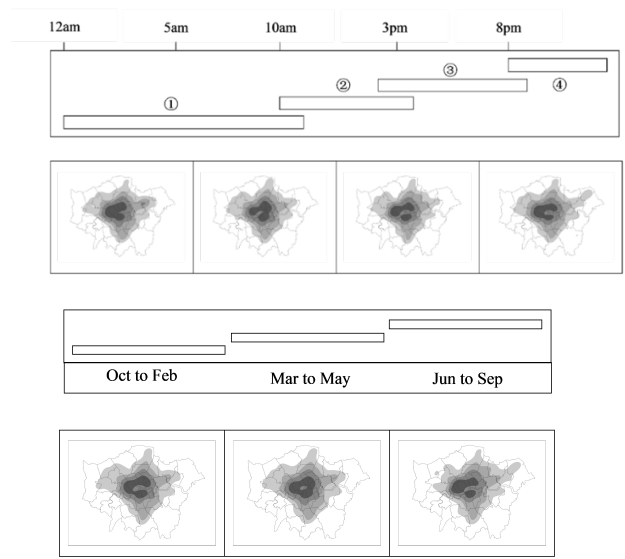

Additionally, the monthly distribution of Flooding incidents showed distinct seasonal patterns. In Figure 21, the highest-density areas for Flooding incidents from June to September were centred around Westminster and Lambeth, while between October and February, these areas expanded and merged. By March to May, the highest-density area continued to spread, though with reduced density at its core. Conversely, the distribution of False Alarm incidents didn't show significant temporal variations.

In general, fluctuations in LFB demands across both temporal and spatial dimensions varied by incident types. Secondary Fire incidents exhibited the most remarkable variations, with broader distribution from June to September and more significant high-density areas from midday to evening. Evident shifts in the highest-demand areas for Flooding incidents within the Special Service category indicated the possibility of resource allocation strategies based on Flooding demand characteristics. In contrast, spatial distribution changed less markedly over time for the primary fire and false alarm demands. While demand volumes of all three incident types showed specific temporal trends throughout the week, differences in spatial distribution between weekends and weekdays were not apparent.

## 5. Discussion

### 5.1. Key Findings

This study systematically analysed the temporal and spatial distribution characteristics of fire (LFB) and ambulance (LAS) services in London, revealing the intersections of service demands across time and space. The key findings are summarised as follows:

(1) Temporal Trend Characteristics

**The two services exhibited high temporal consistency, with peaks in summer and during daytime hours, and were significantly influenced by high temperatures.** In the temporal dimension, LAS and LFB service demands displayed similar cyclical patterns. As shown in Section 4.1.1, both services experienced significant peaks during the summer, with increased demand on Fridays, and daytime peaks between 11:00 and 21:00. Further analysis indicated that both services were similarly correlated with weather variables (temperature, dew point, and wind speed), with temperature showing the strongest positive association. This suggests that extreme heat events likely placed pressure on multiple emergency services.

(2) Spatial Distribution Patterns

**The two service demands overlapped substantially in central areas, with some peripheral zones also emerging as pressure points.** Spatial analysis results demonstrated significant spatial overlap between LFB and LAS service demands within London (Section 4.2.1). Demand hotspots were primarily concentrated in central boroughs such as Westminster, Southwark, and Tower Hamlets, with 8 of the top 10 high-demand boroughs located in Inner London. In addition, limited clusters of high demand were observed in peripheral areas such as Heathrow Airport, northern Croydon, northern Enfield, and Havering. Section 4.2.2 further revealed that, despite differences in the spatial distribution of the three LFB service types, there was a relatively high overlap (61%) between areas showing high values for both LFB and LAS demands, with positive spatial correlations across services. This indicated that certain urban areas consistently faced sustained and intense emergency service pressures. Further examination of dual high-demand areas during non-pandemic periods showed that overall demand levels in Inner London were higher than those in Outer London, with primary peaks concentrated in July and secondary peaks observed in months such as March, May, and October.

(3) Comparison of Driving Factors

**While there were shared drivers, each service demand was influenced by distinct socioeconomic characteristics.** In terms of influencing factors, Section 4.2.3 highlighted that traffic casualties, public transport accessibility, and the proportion of single-person households served as common drivers for both LAS and LFB demands. However, the spatial extent of their influence differed — traffic casualties had a broader impact on LFB demand, while their effect on LAS was more concentrated in South London. Public transport accessibility was more strongly associated with LAS demand along the Thames corridor, whereas its influence on LFB demand was primarily distributed in areas such as Hillingdon and Westminster.

Moreover, other socioeconomic factors displayed distinct associations: LAS demand was significantly correlated with residents' health status, education levels, environmental quality, the proportion of elderly populations, and car-free households. By contrast, LFB demand was more strongly driven by factors such as homeownership rates and crime levels, reflecting its close links with residential patterns and the security environment.

(4) Spatiotemporal Heterogeneity of Different Fire Service Types

**Different types of fire service demands exhibited significant variations across time and space, with Secondary Fires and Flooding incidents particularly standing out.** Analysis in Section 4.3 showed that Secondary Fires displayed clear spatiotemporal seasonal variation: the spatial extent expanded during summer (June–September), with an increase in high-density zones in outer areas; in autumn and winter (October–February), the distribution became more concentrated and less dense. Peak hours occurred from midday to afternoon, with high-density zones expanding north-eastward.

Flooding incidents also showed apparent spatiotemporal patterns: during mornings and afternoons, high-density cores were concentrated around Westminster; during midday, three distinct high-density clusters emerged in the northeastern part of London. Seasonally, high-density areas were concentrated in Westminster and Lambeth from June to September; from October to February, these areas expanded and merged, while from March to May, core density declined but the overall distribution spread further.

### 5.2. Recommendations

Based on the identified shared demand hotspots between the two service agencies, the increasing pressure on emergency services could be addressed through cross-agency collaboration. Such collaboration aimed to integrate existing infrastructure and service resources to ease the response challenges faced by individual agencies during high-demand periods, enhance response speed and effectiveness, and reduce the workload on frontline staff. In light of this, we proposed the following targeted recommendations to further improve the emergency response capacity and resource allocation efficiency in London:

(1) Preparedness for High-Demand Periods

Given the aligned increases in fire and ambulance service demands during summer, on Fridays, and throughout daytime hours (11:00–21:00), and their strong sensitivity to high-temperature weather, it was recommended that cross-agency contingency planning prioritised these high-demand periods. Proactive deployment of personnel and flexible scheduling resources in anticipation of extreme weather events could enhance collaborative response capabilities during peak times.

(2) Targeted Coordination in Dual High-Demand Areas

It was suggested to focus cross-departmental coordination efforts on the dual high-demand areas identified during non-pandemic periods. The top 10 boroughs and associated fire and ambulance stations listed in Section 4.2.1 could serve as priority collaboration nodes. Furthermore, differentiated collaboration strategies between Inner and Outer London were recommended. In Inner London, where demand was more concentrated and seasonal peaks were significant (especially in July, March, May, and October), efforts should focus on building robust mechanisms for highly coordinated responses. In Outer London, targeted coordination mechanisms around key hotspot stations were advised. Although this relatively static deployment strategy differed from the dynamic deployment model currently employed by the LAS, it could nonetheless offer valuable guidance for improving targeted coverage and rapid response.

**Figure 25**
**LAS and LFB station distribution and proposed coordination station**

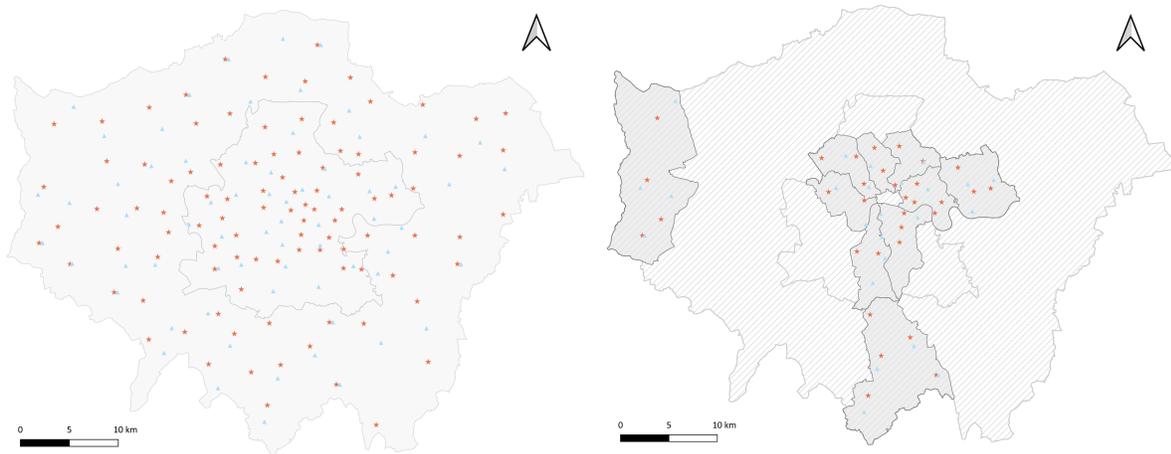

(3) Differentiated Resource Allocation Strategies

Given the observed differences in how socioeconomic characteristics influenced fire and ambulance service demands, it was recommended to incorporate multiple social indicators—such as traffic casualty rates, homeownership rates, and health levels—into the development of tailored response strategies by region and service type. This approach could further enhance the scientific basis and refinement of resource allocation.

(4) Enhanced Dynamic Monitoring and Dispatching for Extreme Events

In light of the strong spatiotemporal fluctuations observed in Secondary Fire and Flooding incidents, it was recommended to establish dynamic monitoring and real-time dispatching mechanisms. Particularly during summer and under extreme weather conditions, efforts should be made to strengthen early warning capabilities and pre-position resources to improve the overall effectiveness of emergency responses.

## 5.3. Limitation and Future Direction

The limitations of this study primarily derived from differences in the resolution of available data. Due to privacy protection restrictions, LAS call data were only available in two forms: monthly statistics at the LSOA level, and hourly statistics aggregated across the entirety of London. This limited the ability to conduct a more detailed analysis of the spatiotemporal dynamics of service demand and affected the comparability with data from the LFB. Future research could benefit from closer collaboration with emergency service agencies to access data with higher spatial and temporal granularity, enabling the identification of demand convergence patterns at finer scales and supporting potential opportunities for cross-agency collaboration.

Moreover, while the present study focused on analysing demand characteristics, it did not incorporate the dynamic evaluation of actual resource deployment. Future work could integrate dynamic ambulance location data and fire dispatch records to construct simulation models—such as agent-based models—that simulate the dynamic response processes of the two service types under different scenarios. This approach would allow the identification of natural convergence features under real-world deployments and further support the development of more practical strategies for cross-agency collaboration.

## 6. Conclusion

As key providers of emergency services in London, the LAS and LFB faced growing pressures from increasing service demand. The UK government has introduced legislation requiring emergency services to strengthen collaboration to enhance overall efficiency and effectiveness. Against this policy background, this study explored the phenomenon of demand overlap between LAS and LFB—two agencies that had traditionally operated independently—across specific temporal and spatial contexts, aiming to provide empirical evidence to support resource deployment optimisation and the promotion of cross-agency collaboration.

This study applied established methods with innovative adaptations, employing bivariate mapping to intuitively visualise the spatial intersections of blue light service demands. The findings revealed significant spatiotemporal synchrony and meaningful, non-random clustering patterns between the two services during specific seasons, time periods, and geographical areas. Based on these characteristics, the study proposed a collaborative approach that strengthened resource coordination and optimised response capabilities within the existing infrastructure and personnel systems. As urban complexity continued to grow and public service budgets remained constrained, the challenge of enhancing the responsiveness and resilience of emergency service systems through more precise demand identification and more efficient collaboration mechanisms has become increasingly critical.

While this study preliminarily validated the potential for cross-agency collaboration and highlighted the enabling role of data-driven approaches in emergency service coordination, it also recognised that this represented only an initial step towards establishing a multi-agency collaborative system. Currently, emergency service agencies still face challenges related to organisational cultural differences, role delineation, and collaboration between strategic and operational levels when jointly addressing shared societal issues. Achieving truly efficient and sustainable collaboration would require not only leveraging the power of data-driven insights and intelligent analytics but also continuously advancing governance framework design, optimising collaboration mechanisms, and building robust evaluation systems to support scientific decision-making and dynamic responsiveness. In conclusion, these findings contribute to the advancement of knowledge in understanding the correlations among emergency services and provide crucial insights for optimizing resource allocation for LAS and LFB, thereby enhancing the efficiency of emergency service responses. This study facilitates a deeper comprehension of the demand patterns for LAS and LFB, supporting more effective planning and resource distribution to address various categories of emergencies.

## Acknowledgement

The author extends appreciation to London Ambulance Service NHS Trust and London Fire Brigade teams for project idea hatching and supports from their expertise.

## Funding Support

This work is not receiving funding support.

## Ethical Statement

This study does not contain any studies with human or animal subjects performed by any of the authors.

## Conflicts of Interest

The authors declare that they have no conflicts of interest to this work.

## Data Availability Statement

The data that support the findings of this study are available on request from the corresponding author, Y Li. The data are not publicly available due to their containing information that could compromise the NDA signed off with NHS Trust.

## Author Contribution Statement

Fangyuan Li[1]: Conceptualization, Methodology, Formal analysis, Resources, Data curation, Writing - origimal draft, Writing - review & editing, Visualization. Yijing Li[2]:  Supervision, Project administration, Writing - review & editing. Luke Edward Rogerson[3]: Resources, Data curation.

## Appendices

Figures and Tables cited in the main text area are as follows.

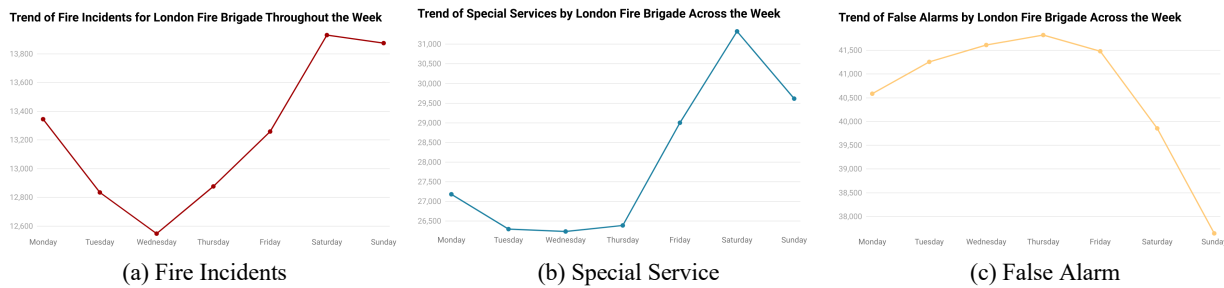

| (a) Fire Incidents | (b) Special Service | (c) False Alarm |

Figure 1. Trends in LFB demand throughout the week for different types

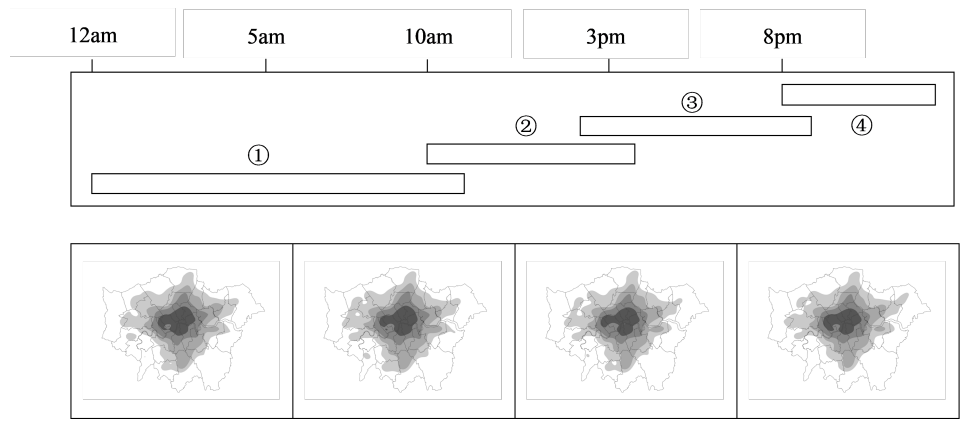

Figure 2. Spatial distribution of special service incidents across different time periods in a day

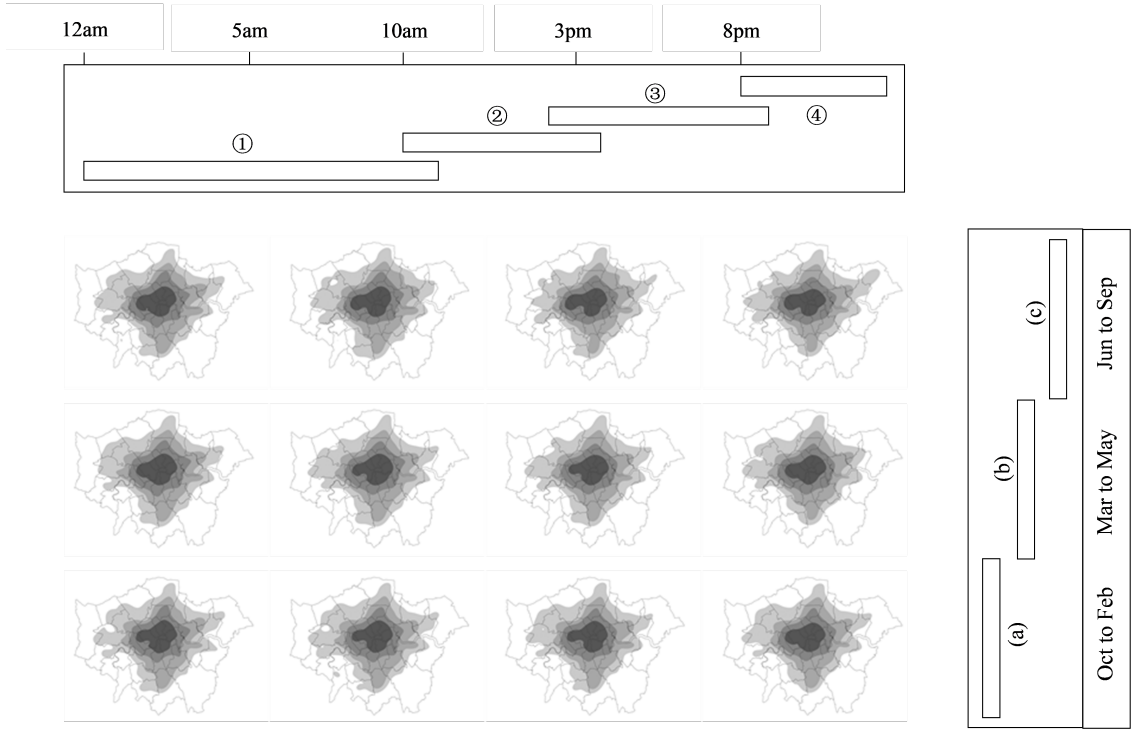

Figure 3. Spatial distribution of special service incidents across different time periods in a day and months in a year

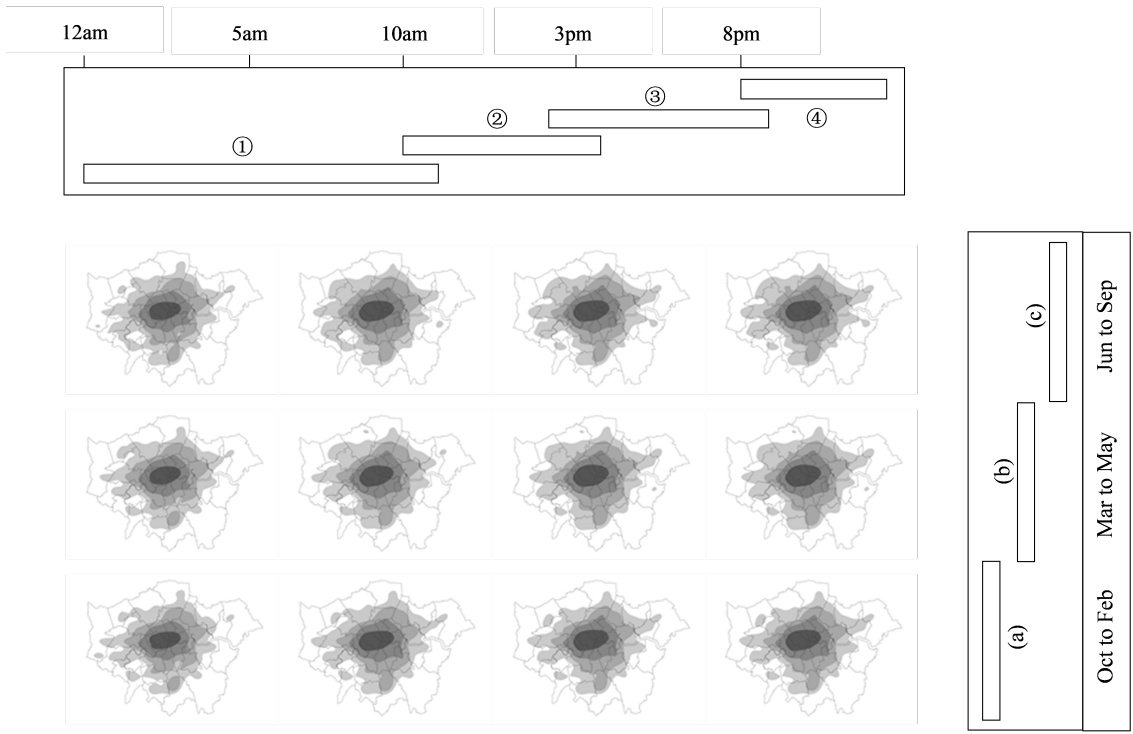

Figure 4. Spatial distribution of false alarm incidents across different time periods in a day and months in a year